\ifx\mnmacrosloaded\undefined \input mn\fi
\def\PsfigVersion{1.10}
\def\setDriver{\DvipsDriver} 
\ifx\undefined\psfig\else \fi
%

\let\LaTeXAtSign=\@
\let\@=\relax
\edef\psfigRestoreAt{\catcode`\@=\number\catcode`@\relax}
\catcode`\@=11\relax
\newwrite\@unused
\def\ps@typeout#1{{\let\protect\string\immediate\write\@unused{#1}}}

\def\DvipsDriver{
	\ps@typeout{psfig/tex \PsfigVersion -dvips}
\def\PsfigSpecials{\DvipsSpecials} 	\def\ps@dir{/}
\def\ps@predir{} }
\def\OzTeXDriver{
	\ps@typeout{psfig/tex \PsfigVersion -oztex}
	\def\PsfigSpecials{\OzTeXSpecials}
	\def\ps@dir{:}
	\def\ps@predir{:}
	\catcode`\^^J=5
}


\def\figurepath{./:}

\def\DoPaths#1{\expandafter\EachPath#1\stoplist}
\def\leer{}
\def\EachPath#1:#2\stoplist{
  \ExistsFile{#1}{\SearchedFile}
  \ifx#2\leer
  \else
    \expandafter\EachPath#2\stoplist
  \fi}
%
%
\def\ps@dir{/}
\def\ExistsFile#1#2{%
   \openin1=\ps@predir#1\ps@dir#2
   \ifeof1
       \closein1
   \else
       \closein1
        \ifx\ps@founddir\leer
           \edef\ps@founddir{#1}
        \fi
   \fi}
%
%
\def\get@dir#1{%
  \def\ps@founddir{}
  \def\SearchedFile{#1}
  \DoPaths\figurepath
}

%
%
\def\@nnil{\@nil}
\def\@empty{}
\def\@psdonoop#1\@@#2#3{}
\def\@psdo#1:=#2\do#3{\edef\@psdotmp{#2}\ifx\@psdotmp\@empty \else
    \expandafter\@psdoloop#2,\@nil,\@nil\@@#1{#3}\fi}
\def\@psdoloop#1,#2,#3\@@#4#5{\def#4{#1}\ifx #4\@nnil \else
       #5\def#4{#2}\ifx #4\@nnil \else#5\@ipsdoloop #3\@@#4{#5}\fi\fi}
\def\@ipsdoloop#1,#2\@@#3#4{\def#3{#1}\ifx #3\@nnil 
       \let\@nextwhile=\@psdonoop \else
      #4\relax\let\@nextwhile=\@ipsdoloop\fi\@nextwhile#2\@@#3{#4}}
\def\@tpsdo#1:=#2\do#3{\xdef\@psdotmp{#2}\ifx\@psdotmp\@empty \else
    \@tpsdoloop#2\@nil\@nil\@@#1{#3}\fi}
\def\@tpsdoloop#1#2\@@#3#4{\def#3{#1}\ifx #3\@nnil 
       \let\@nextwhile=\@psdonoop \else
      #4\relax\let\@nextwhile=\@tpsdoloop\fi\@nextwhile#2\@@#3{#4}}
%
\ifx\undefined\fbox
\newdimen\fboxrule
\newdimen\fboxsep
\newdimen\ps@tempdima
\newbox\ps@tempboxa
\fboxsep = 3pt
\fboxrule = .4pt
\long\def\fbox#1{\leavevmode\setbox\ps@tempboxa\hbox{#1}\ps@tempdima\fboxrule
    \advance\ps@tempdima \fboxsep \advance\ps@tempdima \dp\ps@tempboxa
   \hbox{\lower \ps@tempdima\hbox
  {\vbox{\hrule height \fboxrule
          \hbox{\vrule width \fboxrule \hskip\fboxsep
          \vbox{\vskip\fboxsep \box\ps@tempboxa\vskip\fboxsep}\hskip 
                 \fboxsep\vrule width \fboxrule}
                 \hrule height \fboxrule}}}}
\fi
%
%
\newread\ps@stream
\newif\ifnot@eof       
\newif\if@noisy        
\newif\if@atend        
\newif\if@psfile       
%
%
{\catcode`\%=12\global\gdef\epsf@start{
\def\epsf@PS{PS}
\def\epsf@getbb#1{%
%
%
\openin\ps@stream=\ps@predir#1
\ifeof\ps@stream\ps@typeout{Error, File #1 not found}\else
%
%
   {\not@eoftrue \chardef\other=12
    \def\do##1{\catcode`##1=\other}\dospecials \catcode`\ =10
    \loop
       \if@psfile
	  \read\ps@stream to \epsf@fileline
       \else{
	  \obeyspaces
          \read\ps@stream to \epsf@tmp\global\let\epsf@fileline\epsf@tmp}
       \fi
       \ifeof\ps@stream\not@eoffalse\else
%
%
       \if@psfile\else
       \expandafter\epsf@test\epsf@fileline:. \\%
       \fi
%
%
          \expandafter\epsf@aux\epsf@fileline:. \\%
       \fi
   \ifnot@eof\repeat
   }\closein\ps@stream\fi}%
%
%
\long\def\epsf@test#1#2#3:#4\\{\def\epsf@testit{#1#2}
			\ifx\epsf@testit\epsf@start\else
\ps@typeout{Warning! File does not start with `\epsf@start'.  It may not be a PostScript file.}
			\fi
			\@psfiletrue} 
%
%
{\catcode`\%=12\global\let\epsf@percent=
%
%
%
\long\def\epsf@aux#1#2:#3\\{\ifx#1\epsf@percent
   \def\epsf@testit{#2}\ifx\epsf@testit\epsf@bblit
	\@atendfalse
        \epsf@atend #3 . \\%
	\if@atend	
	   \if@verbose{
		\ps@typeout{psfig: found `(atend)'; continuing search}
	   }\fi
        \else
        \epsf@grab #3 . . . \\%
        \not@eoffalse
        \global\no@bbfalse
        \fi
   \fi\fi}%
%
%
\def\epsf@grab #1 #2 #3 #4 #5\\{%
   \global\def\epsf@llx{#1}\ifx\epsf@llx\empty
      \epsf@grab #2 #3 #4 #5 .\\\else
   \global\def\epsf@lly{#2}%
   \global\def\epsf@urx{#3}\global\def\epsf@ury{#4}\fi}%
%
%
\def\epsf@atendlit{(atend)} 
\def\epsf@atend #1 #2 #3\\{%
   \def\epsf@tmp{#1}\ifx\epsf@tmp\empty
      \epsf@atend #2 #3 .\\\else
   \ifx\epsf@tmp\epsf@atendlit\@atendtrue\fi\fi}


\chardef\psletter = 11 
\chardef\other = 12

\newif \ifdebug 
\newif\ifc@mpute 
\c@mputetrue 

\let\then = \relax
\def\r@dian{pt }
\let\r@dians = \r@dian
\let\dimensionless@nit = \r@dian
\let\dimensionless@nits = \dimensionless@nit
\def\internal@nit{sp }
\let\internal@nits = \internal@nit
\newif\ifstillc@nverging
\def \Mess@ge #1{\ifdebug \then \message {#1} \fi}

{ 
	\catcode `\@ = \psletter
	\gdef \nodimen {\expandafter \n@dimen \the \dimen}
	\gdef \term #1 #2 #3%
	       {\edef \t@ {\the #1}
		\edef \t@@ {\expandafter \n@dimen \the #2\r@dian}%
		\t@rm {\t@} {\t@@} {#3}%
	       }
	\gdef \t@rm #1 #2 #3%
	       {{%
		\count 0 = 0
		\dimen 0 = 1 \dimensionless@nit
		\dimen 2 = #2\relax
		\Mess@ge {Calculating term #1 of \nodimen 2}%
		\loop
		\ifnum	\count 0 < #1
		\then	\advance \count 0 by 1
			\Mess@ge {Iteration \the \count 0 \space}%
			\Multiply \dimen 0 by {\dimen 2}%
			\Mess@ge {After multiplication, term = \nodimen 0}%
			\Divide \dimen 0 by {\count 0}%
			\Mess@ge {After division, term = \nodimen 0}%
		\repeat
		\Mess@ge {Final value for term #1 of 
				\nodimen 2 \space is \nodimen 0}%
		\xdef \Term {#3 = \nodimen 0 \r@dians}%
		\aftergroup \Term
	       }}
	\catcode `\p = \other
	\catcode `\t = \other
	\gdef \n@dimen #1pt{#1} 
}

\def \Divide #1by #2{\divide #1 by #2} 

\def \Multiply #1by #2
       {{
	\count 0 = #1\relax
	\count 2 = #2\relax
	\count 4 = 65536
	\Mess@ge {Before scaling, count 0 = \the \count 0 \space and
			count 2 = \the \count 2}%
	\ifnum	\count 0 > 32767 
	\then	\divide \count 0 by 4
		\divide \count 4 by 4
	\else	\ifnum	\count 0 < -32767
		\then	\divide \count 0 by 4
			\divide \count 4 by 4
		\else
		\fi
	\fi
	\ifnum	\count 2 > 32767 
	\then	\divide \count 2 by 4
		\divide \count 4 by 4
	\else	\ifnum	\count 2 < -32767
		\then	\divide \count 2 by 4
			\divide \count 4 by 4
		\else
		\fi
	\fi
	\multiply \count 0 by \count 2
	\divide \count 0 by \count 4
	\xdef \product {#1 = \the \count 0 \internal@nits}%
	\aftergroup \product
       }}

\def\r@duce{\ifdim\dimen0 > 90\r@dian \then   
		\multiply\dimen0 by -1
		\advance\dimen0 by 180\r@dian
		\r@duce
	    \else \ifdim\dimen0 < -90\r@dian \then  
		\advance\dimen0 by 360\r@dian
		\r@duce
		\fi
	    \fi}

\def\Sine#1%
       {{%
	\dimen 0 = #1 \r@dian
	\r@duce
	\ifdim\dimen0 = -90\r@dian \then
	   \dimen4 = -1\r@dian
	   \c@mputefalse
	\fi
	\ifdim\dimen0 = 90\r@dian \then
	   \dimen4 = 1\r@dian
	   \c@mputefalse
	\fi
	\ifdim\dimen0 = 0\r@dian \then
	   \dimen4 = 0\r@dian
	   \c@mputefalse
	\fi
	\ifc@mpute \then
		\divide\dimen0 by 180
		\dimen0=3.141592654\dimen0
		\dimen 2 = 3.1415926535897963\r@dian 
		\divide\dimen 2 by 2 
		\Mess@ge {Sin: calculating Sin of \nodimen 0}%
		\count 0 = 1 
		\dimen 2 = 1 \r@dian 
		\dimen 4 = 0 \r@dian 
		\loop
			\ifnum	\dimen 2 = 0 
			\then	\stillc@nvergingfalse 
			\else	\stillc@nvergingtrue
			\fi
			\ifstillc@nverging 
			\then	\term {\count 0} {\dimen 0} {\dimen 2}%
				\advance \count 0 by 2
				\count 2 = \count 0
				\divide \count 2 by 2
				\ifodd	\count 2 
				\then	\advance \dimen 4 by \dimen 2
				\else	\advance \dimen 4 by -\dimen 2
				\fi
		\repeat
	\fi		
			\xdef \sine {\nodimen 4}%
       }}

\def\Cosine#1{\ifx\sine\UnDefined\edef\Savesine{\relax}\else
		             \edef\Savesine{\sine}\fi
	{\dimen0=#1\r@dian\advance\dimen0 by 90\r@dian
	 \Sine{\nodimen 0}
	 \xdef\cosine{\sine}
	 \xdef\sine{\Savesine}}}	      

\def\psdraft{
	\def\@psdraft{0}
}
\def\psfull{
	\def\@psdraft{100}
}

\psfull

\newif\if@scalefirst
\def\psscalefirst{\@scalefirsttrue}
\def\psrotatefirst{\@scalefirstfalse}
\psrotatefirst

\newif\if@draftbox
\def\psnodraftbox{
	\@draftboxfalse
}
\def\psdraftbox{
	\@draftboxtrue
}
\@draftboxtrue

\newif\if@prologfile
\newif\if@postlogfile
\def\pssilent{
	\@noisyfalse
}
\def\psnoisy{
	\@noisytrue
}
\psnoisy
\newif\if@bbllx
\newif\if@bblly
\newif\if@bburx
\newif\if@bbury
\newif\if@height
\newif\if@width
\newif\if@rheight
\newif\if@rwidth
\newif\if@angle
\newif\if@clip
\newif\if@verbose
\def\@p@@sclip#1{\@cliptrue}
\newif\if@decmpr
\def\@p@@sfigure#1{\def\@p@sfile{null}\def\@p@sbbfile{null}\@decmprfalse
   \openin1=\ps@predir#1
   \ifeof1
	\closein1
	\get@dir{#1}
	\ifx\ps@founddir\leer
		\openin1=\ps@predir#1.bb
		\ifeof1
			\closein1
			\get@dir{#1.bb}
			\ifx\ps@founddir\leer
				\ps@typeout{Can't find #1 in \figurepath}
			\else
				\@decmprtrue
				\def\@p@sfile{\ps@founddir\ps@dir#1}
				\def\@p@sbbfile{\ps@founddir\ps@dir#1.bb}
			\fi
		\else
			\closein1
			\@decmprtrue
			\def\@p@sfile{#1}
			\def\@p@sbbfile{#1.bb}
		\fi
	\else
		\def\@p@sfile{\ps@founddir\ps@dir#1}
		\def\@p@sbbfile{\ps@founddir\ps@dir#1}
	\fi
   \else
	\closein1
	\def\@p@sfile{#1}
	\def\@p@sbbfile{#1}
   \fi
}
\def\@p@@sfile#1{\@p@@sfigure{#1}}
\def\@p@@sbbllx#1{
		\@bbllxtrue
		\dimen100=#1
		\edef\@p@sbbllx{\number\dimen100}
}
\def\@p@@sbblly#1{
		\@bbllytrue
		\dimen100=#1
		\edef\@p@sbblly{\number\dimen100}
}
\def\@p@@sbburx#1{
		\@bburxtrue
		\dimen100=#1
		\edef\@p@sbburx{\number\dimen100}
}
\def\@p@@sbbury#1{
		\@bburytrue
		\dimen100=#1
		\edef\@p@sbbury{\number\dimen100}
}
\def\@p@@sheight#1{
		\@heighttrue
		\dimen100=#1
   		\edef\@p@sheight{\number\dimen100}
}
\def\@p@@swidth#1{
		\@widthtrue
		\dimen100=#1
		\edef\@p@swidth{\number\dimen100}
}
\def\@p@@srheight#1{
		\@rheighttrue
		\dimen100=#1
		\edef\@p@srheight{\number\dimen100}
}
\def\@p@@srwidth#1{
		\@rwidthtrue
		\dimen100=#1
		\edef\@p@srwidth{\number\dimen100}
}
\def\@p@@sangle#1{
		\@angletrue
		\edef\@p@sangle{#1} 
}
\def\@p@@ssilent#1{ 
		\@verbosefalse
}
\def\@p@@sprolog#1{\@prologfiletrue\def\@prologfileval{#1}}
\def\@p@@spostlog#1{\@postlogfiletrue\def\@postlogfileval{#1}}
\def\@cs@name#1{\csname #1\endcsname}
\def\@setparms#1=#2,{\@cs@name{@p@@s#1}{#2}}
%
%
\def\ps@init@parms{
		\@bbllxfalse \@bbllyfalse
		\@bburxfalse \@bburyfalse
		\@heightfalse \@widthfalse
		\@rheightfalse \@rwidthfalse
		\def\@p@sbbllx{}\def\@p@sbblly{}
		\def\@p@sbburx{}\def\@p@sbbury{}
		\def\@p@sheight{}\def\@p@swidth{}
		\def\@p@srheight{}\def\@p@srwidth{}
		\def\@p@sangle{0}
		\def\@p@sfile{} \def\@p@sbbfile{}
		\def\@p@scost{10}
		\def\@sc{}
		\@prologfilefalse
		\@postlogfilefalse
		\@clipfalse
		\if@noisy
			\@verbosetrue
		\else
			\@verbosefalse
		\fi
}
%
%
\def\parse@ps@parms#1{
	 	\@psdo\@psfiga:=#1\do
		   {\expandafter\@setparms\@psfiga,}}
%
%
\newif\ifno@bb
\def\bb@missing{
	\if@verbose{
		\ps@typeout{psfig: searching \@p@sbbfile \space  for bounding box}
	}\fi
	\no@bbtrue
	\epsf@getbb{\@p@sbbfile}
        \ifno@bb \else \bb@cull\epsf@llx\epsf@lly\epsf@urx\epsf@ury\fi
}	
\def\bb@cull#1#2#3#4{
	\dimen100=#1 bp\edef\@p@sbbllx{\number\dimen100}
	\dimen100=#2 bp\edef\@p@sbblly{\number\dimen100}
	\dimen100=#3 bp\edef\@p@sbburx{\number\dimen100}
	\dimen100=#4 bp\edef\@p@sbbury{\number\dimen100}
	\no@bbfalse
}
\newdimen\p@intvaluex
\newdimen\p@intvaluey
\def\rotate@#1#2{{\dimen0=#1 sp\dimen1=#2 sp
		  \global\p@intvaluex=\cosine\dimen0
		  \dimen3=\sine\dimen1
		  \global\advance\p@intvaluex by -\dimen3
		  \global\p@intvaluey=\sine\dimen0
		  \dimen3=\cosine\dimen1
		  \global\advance\p@intvaluey by \dimen3
		  }}
\def\compute@bb{
		\no@bbfalse
		\if@bbllx \else \no@bbtrue \fi
		\if@bblly \else \no@bbtrue \fi
		\if@bburx \else \no@bbtrue \fi
		\if@bbury \else \no@bbtrue \fi
		\ifno@bb \bb@missing \fi
		\ifno@bb \ps@typeout{FATAL ERROR: no bb supplied or found}
			\no-bb-error
		\fi
		%
%
		\count203=\@p@sbburx
		\count204=\@p@sbbury
		\advance\count203 by -\@p@sbbllx
		\advance\count204 by -\@p@sbblly
		\edef\ps@bbw{\number\count203}
		\edef\ps@bbh{\number\count204}
		\if@angle 
			\Sine{\@p@sangle}\Cosine{\@p@sangle}
	        	{\dimen100=\maxdimen\xdef\r@p@sbbllx{\number\dimen100}
					    \xdef\r@p@sbblly{\number\dimen100}
			                    \xdef\r@p@sbburx{-\number\dimen100}
					    \xdef\r@p@sbbury{-\number\dimen100}}
%
                        \def\minmaxtest{
			   \ifnum\number\p@intvaluex<\r@p@sbbllx
			      \xdef\r@p@sbbllx{\number\p@intvaluex}\fi
			   \ifnum\number\p@intvaluex>\r@p@sbburx
			      \xdef\r@p@sbburx{\number\p@intvaluex}\fi
			   \ifnum\number\p@intvaluey<\r@p@sbblly
			      \xdef\r@p@sbblly{\number\p@intvaluey}\fi
			   \ifnum\number\p@intvaluey>\r@p@sbbury
			      \xdef\r@p@sbbury{\number\p@intvaluey}\fi
			   }
			\rotate@{\@p@sbbllx}{\@p@sbblly}
			\minmaxtest
			\rotate@{\@p@sbbllx}{\@p@sbbury}
			\minmaxtest
			\rotate@{\@p@sbburx}{\@p@sbblly}
			\minmaxtest
			\rotate@{\@p@sbburx}{\@p@sbbury}
			\minmaxtest
			\edef\@p@sbbllx{\r@p@sbbllx}\edef\@p@sbblly{\r@p@sbblly}
			\edef\@p@sbburx{\r@p@sbburx}\edef\@p@sbbury{\r@p@sbbury}
		\fi
		\count203=\@p@sbburx
		\count204=\@p@sbbury
		\advance\count203 by -\@p@sbbllx
		\advance\count204 by -\@p@sbblly
		\edef\@bbw{\number\count203}
		\edef\@bbh{\number\count204}
}
%
%
\def\in@hundreds#1#2#3{\count240=#2 \count241=#3
		     \count100=\count240	
		     \divide\count100 by \count241
		     \count101=\count100
		     \multiply\count101 by \count241
		     \advance\count240 by -\count101
		     \multiply\count240 by 10
		     \count101=\count240	
		     \divide\count101 by \count241
		     \count102=\count101
		     \multiply\count102 by \count241
		     \advance\count240 by -\count102
		     \multiply\count240 by 10
		     \count102=\count240	
		     \divide\count102 by \count241
		     \count200=#1\count205=0
		     \count201=\count200
			\multiply\count201 by \count100
		 	\advance\count205 by \count201
		     \count201=\count200
			\divide\count201 by 10
			\multiply\count201 by \count101
			\advance\count205 by \count201
		     \count201=\count200
			\divide\count201 by 100
			\multiply\count201 by \count102
			\advance\count205 by \count201
		     \edef\@result{\number\count205}
}
\def\compute@wfromh{
		\in@hundreds{\@p@sheight}{\@bbw}{\@bbh}
		\edef\@p@swidth{\@result}
}
\def\compute@hfromw{
	        \in@hundreds{\@p@swidth}{\@bbh}{\@bbw}
		\edef\@p@sheight{\@result}
}
\def\compute@handw{
		\if@height 
			\if@width
			\else
				\compute@wfromh
			\fi
		\else 
			\if@width
				\compute@hfromw
			\else
				\edef\@p@sheight{\@bbh}
				\edef\@p@swidth{\@bbw}
			\fi
		\fi
}
\def\compute@resv{
		\if@rheight \else \edef\@p@srheight{\@p@sheight} \fi
		\if@rwidth \else \edef\@p@srwidth{\@p@swidth} \fi
}
%
\def\compute@sizes{
	\compute@bb
	\if@scalefirst\if@angle
	\if@width
	   \in@hundreds{\@p@swidth}{\@bbw}{\ps@bbw}
	   \edef\@p@swidth{\@result}
	\fi
	\if@height
	   \in@hundreds{\@p@sheight}{\@bbh}{\ps@bbh}
	   \edef\@p@sheight{\@result}
	\fi
	\fi\fi
	\compute@handw
	\compute@resv}
\def\OzTeXSpecials{
	\special{empty.ps /@isp {true} def}
	\special{empty.ps \@p@swidth \space \@p@sheight \space
			\@p@sbbllx \space \@p@sbblly \space
			\@p@sbburx \space \@p@sbbury \space
			startTexFig \space }
	\if@clip{
		\if@verbose{
			\ps@typeout{(clip)}
		}\fi
		\special{empty.ps doclip \space }
	}\fi
	\if@angle{
		\if@verbose{
			\ps@typeout{(rotate)}
		}\fi
		\special {empty.ps \@p@sangle \space rotate \space} 
	}\fi
	\if@prologfile
	    \special{\@prologfileval \space } \fi
	\if@decmpr{
		\if@verbose{
			\ps@typeout{psfig: Compression not available
			in OzTeX version \space }
		}\fi
	}\else{
		\if@verbose{
			\ps@typeout{psfig: including \@p@sfile \space }
		}\fi
		\special{epsf=\ps@predir\@p@sfile \space }
	}\fi
	\if@postlogfile
	    \special{\@postlogfileval \space } \fi
	\special{empty.ps /@isp {false} def}
}
\def\DvipsSpecials{
	\special{ps::[begin] 	\@p@swidth \space \@p@sheight \space
			\@p@sbbllx \space \@p@sbblly \space
			\@p@sbburx \space \@p@sbbury \space
			startTexFig \space }
	\if@clip{
		\if@verbose{
			\ps@typeout{(clip)}
		}\fi
		\special{ps:: doclip \space }
	}\fi
	\if@angle
		\if@verbose{
			\ps@typeout{(clip)}
		}\fi
		\special {ps:: \@p@sangle \space rotate \space} 
	\fi
	\if@prologfile
	    \special{ps: plotfile \@prologfileval \space } \fi
	\if@decmpr{
		\if@verbose{
			\ps@typeout{psfig: including \@p@sfile.Z \space }
		}\fi
		\special{ps: plotfile "`zcat \@p@sfile.Z" \space }
	}\else{
		\if@verbose{
			\ps@typeout{psfig: including \@p@sfile \space }
		}\fi
		\special{ps: plotfile \@p@sfile \space }
	}\fi
	\if@postlogfile
	    \special{ps: plotfile \@postlogfileval \space } \fi
	\special{ps::[end] endTexFig \space }
}
%
%
\def\psfig#1{\vbox {
	%
	\ps@init@parms
	\parse@ps@parms{#1}
	\compute@sizes
	\ifnum\@p@scost<\@psdraft{
		\PsfigSpecials 
		\vbox to \@p@srheight sp{
			\hbox to \@p@srwidth sp{
				\hss
			}
		\vss
		}
	}\else{
		\if@draftbox{		
			\hbox{\fbox{\vbox to \@p@srheight sp{
			\vss
			\hbox to \@p@srwidth sp{ \hss 
			 \hss }
			\vss
			}}}
		}\else{
			\vbox to \@p@srheight sp{
			\vss
			\hbox to \@p@srwidth sp{\hss}
			\vss
			}
		}\fi

	}\fi
}}
\psfigRestoreAt
\setDriver
\let\@=\LaTeXAtSign

\long\def\crap#1{}
\def\d{{\rm d}}\def\p{\partial}
\def\E{{\cal E}}
\def\L{{\cal L}}
\def\P{{\cal P}}
\def\pr{\mathop{\smash{\rm p}\vphantom{\sin}}}
\def\b#1{{\bmath#1}}
\def\=#1{\overline{#1}}

\def\lta{\la}\def\gta{\ga}

\def\km{{\rm\,km}}
\def\kms{{\rm\,km\,s^{-1}}}
\def\kpc{{\rm\,kpc}}
\def\Mpc{{\rm\,Mpc}}
\def\msun{{\rm\,M_\odot}}
\def\lsun{{\rm\,L_\odot}}
\def\rsun{{\rm\,R_\odot}}
\def\pc{{\rm\,pc}}
\def\cm{{\rm\,cm}}
\def\yr{{\rm\,yr}}
\def\Gyr{{\rm\,Gyr}}
\def\Myr{{\rm\,Myr}}
\def\au{{\rm\,AU}}
\def\gm{{\rm\,g}}
\def\kg{{\rm\,kg}}
\def\ergps{{\rm\,erg\,s}^{-1}}
\def\K{{\rm\,K}}
\def\rms{{\caps rms}}

\def\eqname#1{%
   \global\advance\Eqnno by1
   \xdef#1{(\theeq)}
   \global\advance\Eqnno by-1
}
\everydisplay{\puteqnum}
\def\puteqnum#1$${#1\eqno\stepeq$$}
\def\refeq#1{%
 \advance\Eqnno by -#1
 \advance\Eqnno by 1
 \hbox{(\theeq)}%
 \advance\Eqnno by #1
 \advance\Eqnno by-1
}

\def\ifundefined#1{%
   \expandafter\ifx\csname#1\endcsname\relax
}
\def\ref#1{\ifundefined{#1} $\bullet$#1$\bullet$%
  \immediate\write16{Undefined reference #1}%
            \write-1{Undefined reference #1}
\else\hbox{\csname#1\endcsname}\fi
}

\newif\ifAMStwofonts

\ifCUPmtplainloaded \else
  \NewTextAlphabet{textbfit} {cmbxti10} {}
  \NewTextAlphabet{textbfss} {cmssbx10} {}
  \NewMathAlphabet{mathbfit} {cmbxti10} {} 
  \NewMathAlphabet{mathbfss} {cmssbx10} {} 
  \ifAMStwofonts
    \NewSymbolFont{upmath} {eurm10}
    \NewSymbolFont{AMSa} {msam10}
    \NewMathSymbol{\upi}     {0}{upmath}{19}
    \NewMathSymbol{\umu}     {0}{upmath}{16}
    \NewMathSymbol{\upartial}{0}{upmath}{40}
    \NewMathSymbol{\leqslant}{3}{AMSa}{36}
    \NewMathSymbol{\geqslant}{3}{AMSa}{3E}
    \let\oldle=\le     \let\oldleq=\leq
    \let\oldge=\ge     \let\oldgeq=\geq
    \let\leq=\leqslant \let\le=\leqslant
    \let\geq=\geqslant \let\ge=\geqslant
  \else
    \def\umu{\mu}
    \def\upi{\pi}
    \def\upartial{\partial}
  \fi
\fi

\pageoffset{-2.5pc}{0pc}

\loadboldmathnames




\begintopmatter  

\title{Mass profiles and anisotropies of early-type galaxies}
\author{John Magorrian and David Ballantyne}
\affiliation{CITA, University of Toronto, 60 St George Street,
  Toronto, Ontario, Canada M5S 3H8, and}
\vskip0.1truecm
\affiliation{Institute of Astronomy, Madingley Road, Cambridge
CB3 0HA}

\shortauthor{S. J. Magorrian and D. R. Ballantyne}
\shorttitle{Mass profiles and anisotropies of early-type galaxies}


\abstract{%
We discuss the problem of using stellar kinematics of early-type
galaxies to constrain the galaxies' orbital anisotropies and radial
mass profiles.
We demonstrate that compressing a galaxy's light distribution along
the line of sight produces approximately the same signature in the
line-of-sight velocity profiles as radial anisotropy.  In particular,
fitting spherically symmetric dynamical models to apparently round,
isotropic face-on flattened galaxies leads to a spurious bias towards
radial orbits in the models, especially if the galaxy has a weak
face-on stellar disk.  Such face-on stellar disks could plausibly be
the cause of the radial anisotropy found in spherical models of
intermediate luminosity ellipticals such as NGC~2434, NGC~3379 and
NGC~6703.

In the light of this result, we use simple dynamical models to
constrain the outer mass profiles of a sample of 18 round, early-type
galaxies.  The galaxies follow a Tully--Fisher relation parallel to
that for spiral galaxies, but fainter by at least 0.8 mag ($I$-band)
for a given mass.  The most luminous galaxies show clear evidence for
the presence of a massive dark halo, but the case for dark haloes in
fainter galaxies is more ambiguous.  We discuss the observations that
would be required to resolve this ambiguity.}

\keywords {celestial mechanics -- stellar dynamics
           -- galaxies: kinematics and dynamics
           -- galaxies: halos}

\maketitle  

\section {Introduction}

While it has long been known that spiral galaxies are embedded in
halos of some kind of dark component, the evidence that ellipticals
have dark halos is much more patchy.  Analyses of the temperature
profiles (e.g., Brighenti \& Mathews~1997), the mean temperature
(Loewenstein \& White 1999) and the shapes (Buote \& Canizares~1997)
of the hot, X-ray emitting gas in bright ellipticals all demonstrate
that these large galaxies have dark halos extending to many effective
radii.  Similarly, results from gravitational lensing studies (e.g.,
Kochanek~1996) show that the mass-to-light ratios of at least some
large ellipticals must rise beyond an effective radius or so.  It was
unclear whether faint, compact ellipticals also had dark halos until
the recent work of Rix et al.\ (1997) and Gerhard et al.\ (1998).
These authors applied sophisticated spherical dynamical models to deep
measurements of the line-of-sight velocity profiles (VPs) of the round
galaxies NGC~2434 and NGC~6703.  They found that the mass-to-light
ratio in each rises outwards beyond about an effective radius.  But
despite this recent progress, our understanding of the properties of
dark haloes of ellipticals remains poor.

The main difficulty in determining the mass distributions (or,
equivalently, the potentials) of most ellipticals is the absence of
dynamical tracers other than stars, planetary nebulae (PNe) and
globular clusters.  Since these tracers are collisionless, one cannot
determine the galaxy's potential without simultaneously constraining
the distribution of orbits of the tracer population (the orbit
distribution function, or DF) within the galaxy.  There are two
general approaches to this problem.  One is to consider a range of
parametrized trial galaxy potentials and, for each trial potential, to
search for the DF that best matches the observed VPs.  Potentials for
which no satisfactory fit can be found can be ruled out.  This is the
approach used by Merritt \& Saha (1993), Merritt (1993), Rix et al.\
(1997) and Gerhard et al.\ (1998).  An alternative approach is to
consider a range of simple parametrized forms for the DF and to search
for the mass distributions consistent with the assumed DFs and the
available observations.  This has been used by Merritt \& Tremblay
(1993), Gebhardt \& Fischer (1995) and Merritt (1996), but only
considering one form for the DF (isotropy) and without making full use
of the information contained in the observed VPs.  We follow this
second approach in the present paper, but using a wider range of
parametrized DFs and making fuller use of observed VPs.

The paper is organized as follows.  In common with most other work in
this area, our models assume spherical symmetry for convenience.  So,
we begin in \S2 by identifying the biases that this assumption
introduces.  In \S3 we consider the simple problem of determining, for
a spherical galaxy, the range of mass-to-light ratio profiles
consistent with an observed velocity dispersion profile and some
assumed form for the anisotropy of the velocity dispersion tensor.
Section~4 then shows how this method can naturally be extended to use
VP shape information, including information from discrete tracers of
the VPs, such as PNe.  Applications to real galaxies follow in \S5,
before we sum up in \S6.

\section{The degeneracy between mass, anisotropy and flattening}

Throughout this paper we will assume that our round galaxies are close
to spherical.  Before plunging into the details of the modelling
procedure, let us first pause to consider the consequences of this
assumption.  It is instructive to take a real galaxy as an example.
We use NGC~2434, an E0 studied in detail by Rix et al.\ (1997).

\beginfigure{1}
\centerline{\psfig{file=n2434kin.ps,width=\hsize}}
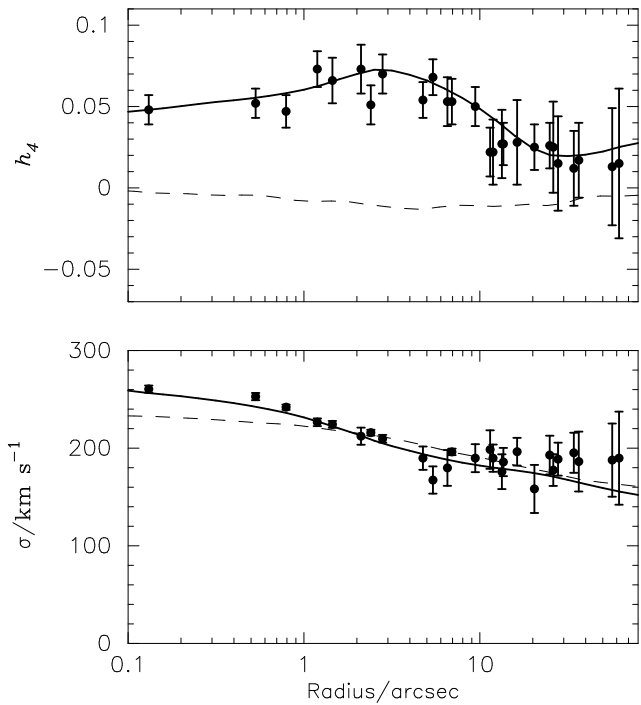
\caption{{\bf Figure 1(a).}  Kinematics of the E0 galaxy NGC~2434.
The points on plot the observed velocity dispersion~$\sigma$ and
Gauss--Hermite VP shape parameter $h_4$ versus radius.  The broken
curves plot the corresponding quantities for spherical isotropic
galaxy models.  The heavy solid curves show the kinematics of an
flattened isotropic model of the galaxy, viewed face-on.  Figure~1(b)
shows the edge-on surface brightness distribution of this model.}
\endfigure
\beginfigure{2}
\centerline{
            \psfig{file=n2434iso.ps,width=\hsize}}
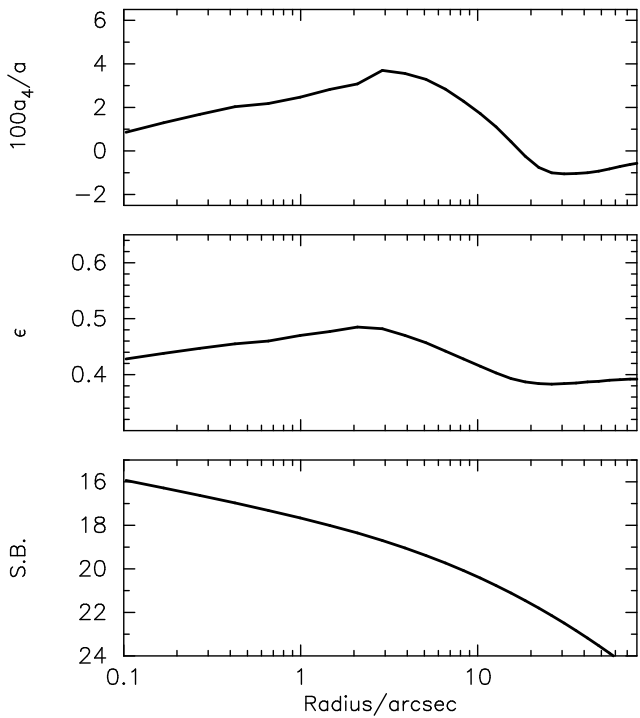
\caption{{\bf Figure 1(b).}  Edge-on surface brightness distribution of the
flattened isotropic model in Fig.~1(a).  The panels show the model's
isophotal shape parameters (surface brightness, ellipticity and
lowest-order shape parameter $a_4/a$), using the convention of Bender
\& M\"ollenhoff (1987).}
\endfigure

The points on Figure~1(a) plot the observed stellar kinematics of this
galaxy.  In addition to the usual velocity dispersion~$\sigma$, there
is a Gauss--Hermite parameter $h_4$ that measures the deviation of the
observed VP from the underlying best-fit Gaussian (van der Marel \&
Franx~1993).  VPs more boxy than Gaussian have $h_4<0$, whereas those
more triangular have $h_4>0$.  The dashed curves plot the kinematics
of a spherical model galaxy with the same projected light distribution
as NGC~2434, an isotropic velocity distribution, and an isothermal
potential with $v_{\rm c}=300\kms$.  The model reproduces the observed
velocity dispersion quite well, but not the shapes of the VPs -- its
VPs are not centrally peaked enough.  This can be rectified by
increasing the fraction of stars on radial orbits in the model, while
keeping the light distribution and potential fixed.  Indeed, Rix et
al.\ (1997) have used sophisticated modelling machinery based on
Schwarzschild's (1979) method to construct a strongly radially
anisotropic spherical model that provides an excellent fit to the
observed kinematics of this galaxy.

It turns out, however, that there is no need to invoke radial
anisotropy to explain these kinematics.  Let us start with the
spherical isotropic model again, but, instead of increasing the
fraction of stars on radial orbits, let us keep the velocity
distribution isotropic while flattening the model along the line of
sight.
Flattening the model like this increases the fraction of stars
orbiting in the plane of the sky (i.e., orbits with high $L_z$) and
decreases the model's projected velocity dispersion.  This was first
noted by Bender, Saglia \& Gerhard (1994), who invoked almost face-on
circular orbits in a similar way to explain the minor-axis VPs of
NGC~4660.  The extra orbits in the plane of the sky affect the VP
shapes in the same way as a bias towards radial orbits in a spherical
galaxy.

To give a more concrete example, we have increased the mass of our
spherical isotropic model of NGC~2434 from $v_{\rm c}=300\kms$ to
$400\kms$, and squeezed its light distribution by first making it
oblate with axis ratio $q=0.6$ and then adjusting its $\cos4\theta$
Fourier coefficients until its edge-on projected surface brightness
distribution has the form shown in Figure~1(b).  The $a_4$ coefficient
on this plot measures the lowest-order symmetric deviation of each
isophote from the underlying best-fit ellipse.  The model is quite
disky ($a_4>0$), with a peak $a_4/a$ of 3.7\% when viewed edge-on, or
1.9\% when averaged over all viewing angles.

We calculate the kinematics of the model using the method described in
Magorrian \& Binney (1994).  The results are plotted as the heavy
curves on Figure~1(a).  This flattened, isotropic model provides just
as good a fit to the observations as the radially anisotropic,
spherical models of Rix et al.\ (1997).
While the diskiness of our model affects the shape of its VPs when
seen face-on, even when seen edge on the disk does not have a strong
effect on the model's $(v_0/\sigma)$, the characteristic ratio of the
mean-streaming velocity to the velocity dispersion (Binney~1978;
Davies et al.\ 1983).  This remains very close to the predictions for
an oblate isotropic rotator.

Of course, these results do not mean that NGC~2434 {\it must} be
isotropic with a weak face-on stellar disk, but they do show that
making unwarranted assumptions about a galaxy's intrinsic shape will
lead to biases in the orbit distribution that one finds.  Throughout
the rest of this paper we will model our round galaxies under the
assumption of spherical symmetry.  In Appendix~A we quantify the
degeneracy between mass, anisotropy and shape, showing how the results
from our spherical models may be corrected for any supposed degree of
(constant) flattening.

\section {Constraining the mass profile using velocity dispersion measurements}

We are given measurements of a galaxy's projected velocity dispersion
$\sigma_i$ with corresponding errors (assumed Gaussian)
$\Delta\sigma_i$ at points $R_i$, with $i=1\ldots N$.  The luminosity
density of the galaxy is $j(r)$.  We seek the range of smooth
mass-to-light ratio profiles $\Upsilon(r)$ consistent with these
observations and some assumed form for $\beta(r)$, the anisotropy of
the galaxy's velocity dispersion tensor, where
$$\beta(r)\equiv1-{\sigma_\phi^2(r)\over\sigma_r^2(r)},
$$
and $\sigma_r(r)$ and $\sigma_\phi(r)$ are the velocity dispersions in
the radial and azimuthal directions at radius~$r$.  Before explaining
how we determine the range of $\Upsilon(r)$, we first describe the
necessary simpler step in the method: how to calculate the predicted
velocity dispersion profile $\hat\sigma(R)$ of a trial $\Upsilon(r)$.

\subsection{From $\Upsilon(r)$ to $\hat\sigma(R)$}

Given $\Upsilon(r)$, the mass interior to radius~$r$ is given by
$$M(r)=4\pi\int_0^r\Upsilon(r')j(r')r'^2\,\d r'.
\eqname\intone
$$
The Jeans equation (e.g., Binney \& Tremaine~1987),
$${\d (j\={v_r^2})\over\d r} + 2\beta{j\={v_r^2}\over r}
=-j{GM\over r^2},
\eqname\jeans
$$
can be integrated to give the intrinsic second-order moments in terms
of $j(r)$, $M(r)$ and $\beta(r)$:
$$j\={v_r^2}(r)=
\int_r^\infty\d r'\, \nu(r'){GM(r')\over r'^2}
\exp\left[\int_r^{r'}{2\beta(r'')\over r''}\d r''\right],
\eqname\inttwo
$$
and, by definition,
$$\={v_\phi^2}=\={v_\theta^2}=(1-\beta)\={v_r^2}.
$$
Projecting along lines of sight, the zeroth and second-order moments
at projected radius~$R$ are
$$\eqalign{
I(R) &
   = 2\int_R^\infty{j(r) r\,\d r\over \sqrt{r^2-R^2}}\cr
I\={v_{\rm p}^2}(R) &
   = 2\int_R^\infty{j\={v_r^2}(r) r\,\d r\over \sqrt{r^2-R^2}} 
        \left(1-\beta(r){R^2\over r^2}\right).\cr
}
\eqname\intthree
$$
To include the effects of seeing, we assume that the PSF of the
kinematic observations is a Gaussian with dispersion
$\sigma_\star={\rm FWHM}/2.305$,
$${\rm PSF}(\Delta x)={1\over2\pi\sigma_\star^2}
\exp\left[-{1\over2}\left(\Delta x\over\sigma_\star\right)^2\right].
$$
The convolution of any projected distribution~$\mu_0(R)$ (e.g., $I(R)$
or $I\={v_{\rm p}^2}(R)$) with this PSF is
$$\eqalign{
\mu(R) & = 
{1\over\sigma_\star^2}\exp\left(-{R^2\over2\sigma_\star^2}\right)\times\cr
& \qquad\int_0^\infty R'\exp\left(-{R'^2\over2\sigma_\star^2}\right)
I_0\left(RR'\over\sigma_\star^2\right)\mu_0(R')\,\d R',\cr }
\eqname\intfour
$$
where $I_0(x)$ is the zeroth-order modified Bessel function.  Dividing
the seeing-convolved second-order moment by the zeroth order one and
taking the square root yields the model's predicted $\hat\sigma(R)$,
and, in particular, its predictions $\hat\sigma_i$ at the observed
points $R_i$.

In the numerical implementation of the calculations above we store the
logarithms of $j(r)$, $M(r)$, etc. on grids spaced logarithmically in
radius from the radius of the innermost observed isophote to a few
times the radius of the outermost one, typically with about 30 grid
points.  Values of $j(r)$, $M(r)$, etc. at intermediate points are
obtained by linear interpolation on these log--log grids and
exponentiating.  Since the positions of the radial grid points, as
well as $j(r)$ and $\beta(r)$, are independent of $\Upsilon(r)$, many
of the coefficients involved in the numerical calculation of the
integrals \ref{intone}, \ref{inttwo},
\ref{intthree} and \ref{intfour} need only be evaluated once, leading
to a considerable decrease in the time needed to calculate the
$\hat\sigma_i$ from a trial $\Upsilon(r)$.

\subsection {Determining the range of acceptable $\Upsilon(r)$}

By Bayes' theorem, the probability of any model $\Upsilon(r)$ given
the available velocity dispersion data $D$ and our assumed $\beta$
satisfies
$$\P\equiv\pr(\Upsilon|D,\beta)\propto \pr(D|\Upsilon,\beta)\pr(\Upsilon).
\eqname\posterior
$$
Since we assume that the observational errors are Gaussian, the
likelihood $\pr(D|\Upsilon,\beta)$ is given by
$$\pr(D|\Upsilon,\beta)=\exp\left(-{1\over2}\chi^2\right),
\eqname\likesig
$$
where
$$\chi^2[\Upsilon,\beta]=\sum_{i=1}^{N}
\left(\hat\sigma_i-\sigma_i\over\Delta\sigma_i\right)^2.
\eqname\chisqsig
$$
The prior $\pr(\Upsilon)$ encodes our prejudices about what
constitutes a reasonable $\Upsilon(r)$.  Our only expectation is that
it be smooth, and we take
$$\ln\pr(\Upsilon)=-\lambda
\int\left(\d^2\ln\Upsilon\over\d(\ln r)^2\right)^2\d\ln r
\cdot\int\d\ln r,
\eqname\prior
$$
which uses the mean-square change in the slope of $\ln\Upsilon$ versus
$\ln r$ as a measure of smoothness.  For a reasonable $\Upsilon(r)$
the change in the mean-square slope would be less than about 100.  We
set $\lambda=1/100$ so that an increase in the mean-square slope by
100 has the same effect on $\pr(\Upsilon|D,\beta)$ as an increase in
$\chi^2$ of one.

To obtain the range of $\Upsilon(r)$ that maximize $\P$, we first fit
an initial guess of the form
$$\Upsilon(r)=\Upsilon_0\left(r\over r_0\right)^a
\left(1+{r\over r_0}\right)^b,
\eqname\inifit
$$
using the downhill simplex method (Press et al.\ 1992) to find the
parameters $(\Upsilon_0,r_0,a,b)$ that minimize $\chi^2$ in
equation~\ref{chisqsig}.  Then the Metropolis algorithm (Metropolis et
al.\ 1953) is used to explore the range of $\Upsilon(r)$ consistent
with the data:
\begingroup\parindent=3em
\item{(1)} Calculate $\P[\Upsilon]$ for the initial guess.

\item{(2)} Make a change $\delta\Upsilon$ to $\Upsilon$.  Calculate
$\chi^2[\Upsilon+\delta\Upsilon]$, $\pr(\Upsilon+\delta\Upsilon)$
and thus $\P[\Upsilon+\delta\Upsilon]$.

\item{(3)} if $\Delta\P\equiv
\P[\Upsilon+\delta\Upsilon]-\P[\Upsilon]>0$ accept the change
$\delta\Upsilon$; otherwise accept it with probability
$\exp(\Delta\P)$.

\item{(4)} Go back to step 2.

\endgroup
Each change $\delta\Upsilon$ in step (2) is made by choosing a single
radial grid point~$i$ at random and adding $x\Delta_i$ to
$\ln\Upsilon_i$, where $x$ is a random number uniformly distributed in
$(-1,1)$ and the constant $\Delta_i$ sets the size of the maximum
change in $\ln\Upsilon$ at the point~$i$.  For maximum efficiency the
$\Delta_i$ should be chosen so that the change in each step of the
Metropolis algorithm has roughly equal chances of being accepted or
rejected.  We ensure this is the case by initially choosing all
$\Delta_i=1$ and following the Metropolis algorithm from the initial
fit~\ref{inifit} for enough iterations to ensure that each radial grid
point is sampled many times.  During this time, we increase $\Delta_i$
by a small factor (e.g., 1.5) if a change in the $i^{\rm th}$ grid
point is accepted, and decrease it by the same factor if the change is
rejected.  We then restart the Metropolis algorithm from the initial
fit~\ref{inifit} keeping the $\Delta_i$ fixed.

As the Metropolis algorithm progresses, the probability of any given
configuration $\Upsilon(r)$ occurring tends towards the probability
$\pr(\Upsilon|D,\beta)\equiv\P$.  This makes it easy to obtain
confidence intervals on $\Upsilon(r)$ and other model quantities.  In
addition, if we have two models, one with anisotropy $\beta_1(r)$, the
other with $\beta_2(r)$, and no reason for favouring one over the
other ($\pr(\beta_1)=\pr(\beta_2)$), then a simple application of
Bayes' theorem yields their relative probability,
$${\pr(\beta_1|D)\over\pr(\beta_2|D)}={\pr(D|\beta_1)\over\pr(D|\beta_2)}=
{\int\pr(D|\Upsilon,\beta_1)\pr(\Upsilon)\,\d\Upsilon\over
\int\pr(D|\Upsilon,\beta_2)\pr(\Upsilon)\,\d\Upsilon},
\eqname\comparemodel
$$
which from equation~\ref{posterior} is just the ratio of the average
values of $\P$ in the Metropolis algorithm.  Of course,
equation~\refeq1 is not very useful in the present case, since for
most choices of $\beta(r)$ there is usually some smooth $\Upsilon(r)$
that projects to a good fit to the observed $\sigma_i$.  But given
further information on the shapes of the VPs (Section~4), and if we
somehow knew that the galaxy was close to spherical, we could use
equation~\refeq1 to distinguish between models with different
$\beta(r)$.

\subsection{An application to M87}

In principle, the mass profile can also be obtained by rearranging the
Jeans equation~\ref{jeans} to get
$$M(r)=-{r\={v_r^2}\over G}
\left({\d\ln j\over\d\ln r}+{\d\ln\={v_r^2}\over\d\ln r}+2\beta\right),
\eqname\encmass
$$ 
in which $\={v_r^2}$ follows from an inversion of
equation~\ref{intthree}.  This approach was used by Sargent et al.\
(1978) to obtain the mass distribution of M87 assuming that it is
isotropic ($\beta=0$).  The method is very direct, but requires the
fitting of a smooth, arbitrary $\sigma(R)$ profile to the observed
$\sigma_i$, leading to complex biases in the results.  Here we use
Sargent et al.'s data and results to provide a simple test of our
method.

\beginfigure{3}
\centerline{\psfig{file=sm87.ps,width=\hsize}}
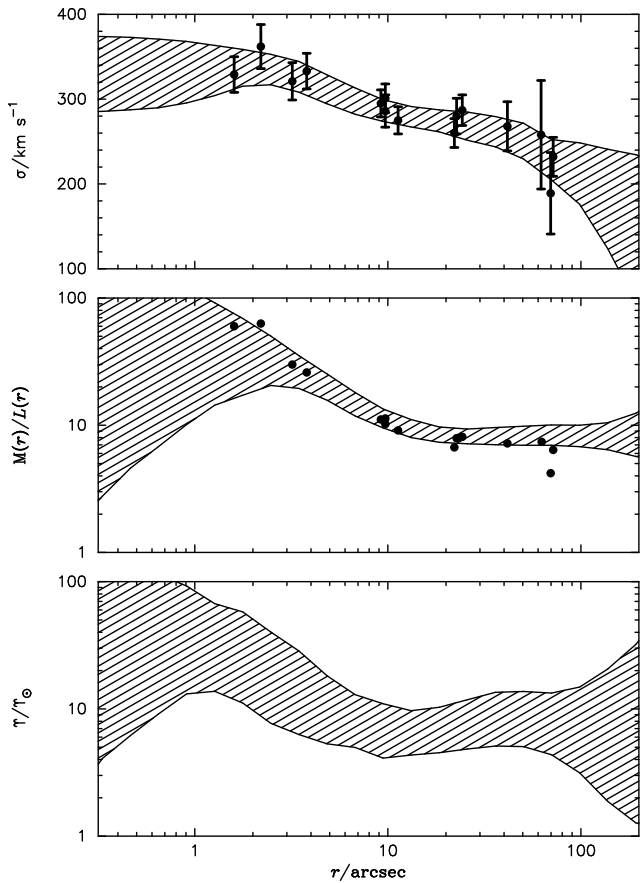
\caption{{\bf Figure 2.} Our model for the mass distribution of M87,
using the data of Sargent et al.\ (1978) and assuming the galaxy is
isotropic ($\beta=0$).  The bottom panel shows the 95\% confidence
limits on the the local mass-to-light ratio $\Upsilon(r)$.  The
corresponding limits on the ``global'' mass-to-light ratio $M(r)/L(r)$
and the model's projected velocity dispersions are plotted on the
panels above.  For comparison, the points show the $M(r)/L(r)$
obtained by Sargent et al.\ (middle panel) and the observed
dispersions (top panel). }
\endfigure

We obtain our model luminosity density $j(r)$ from their photometry
using the method described in Magorrian (1999).  It is unclear exactly
what seeing was in effect at the time of Sargent et al.'s
observations, but we assume 2 arcsec FWHM.  We then apply the
algorithm above to their kinematical data for $10^5$ iterations, the
results of which are plotted on Fig.~2.  The panels show the 95\%
confidence limits on the local mass-to-light ratio $\Upsilon(r)$, the
``average'' mass-to-light ratio $M(r)/L(r)$ enclosed within
radius~$r$, and the corresponding limits on the model's projected
dispersion profile $\hat\sigma(r)$, together with the observed
$\sigma_i$.  The values of $M(r)/L(r)$ found by Sargent et al.\ are
also plotted on Fig.~2 for comparison with ours.  There is quite good
agreement between our results and theirs, except within the innermost
few arcsec of the galaxy centre.  The reason for this is simple (e.g.,
Binney \& Tremaine~1987): at large radii, the $j(r)$ profile is steep
($\sim r^{-3}$) while the $\={v_r^2}$ profile is shallow ($\sim r^0$),
so that the value of the bracketed expression in~\refeq1 used by
Sargent et al.\ is dominated by the relatively well-determined slope
of $j(r)$.  At smaller radii, $j(r)$ becomes much shallower,
$\={v_r^2}$ steepens, and so $M(r)$ is affected more by the poorly
determined slope of the latter.

While the agreement between the results from the two methods is
gratifying, it is important to note the fundamental difference between
them.  In methods like Sargent et al.'s (e.g., Gebhardt \&
Fischer~1995), one is required to make assumptions about unobserved
data, whereas in methods like ours (e.g., Merritt \& Tremblay~1993;
Merritt~1996) one uses only real, observed data, and confines any
assumptions to the model itself.  In the former case, there is no
straightforward way to obtain confidence bounds on the model's mass
density profiles, or indeed to impose the reasonable constraint that
the mass density must be non-negative.  On the other hand, our model's
disregard for imaginary data is why the exact amount of mass enclosed
within the innermost few arcsec is quite uncertain -- due to the
effects of seeing, the data are consistent with quite a gentle
increase in mass-to-light ratio near the centre.  Similarly, our model
correctly reproduces the fact that there are no interesting
constraints on $\Upsilon(r)$ beyond the radius of the outermost
observed dispersion.  Of course, the one big advantage of methods like
Sargent et al.'s over ours is that they are relatively quick, and can
be adequate if one is content with finding only a restricted range of
the full set of plausible mass density profiles.

\def\ra{r_{\rm a}}\def\vp{v_{\rm p}}
\def\l{l}\def\hermone_#1{{H_{\rm e}}_#1}
\def\wgc{w_{\rm g}}\def\sigmagc{\sigma_{\rm g}}
\def\wgh{w_{\rm g}}
\def\gammagh{\gamma_{\rm g}}\def\sigmagh{\sigma_{\rm g}}

\section {Using velocity profiles to constrain both the mass
profile and the anisotropy}

The method above can be extended to make use of the full observed VPs
if, instead of assuming a form for $\beta(r)$, we make a stronger
assumption about the form of the galaxy's distribution function (DF).
By Jeans' theorem, the DF of a spherical galaxy is a function only of
the stars' binding energies $\E$ and angular momenta $L$ per unit
mass.  We make the following ansatz for the DF (Cuddeford~1991):
$$f(\E,L^2)=L^{-2\beta_0}f_0(Q)\qquad
\hbox{with\ } Q\equiv \E-{L^2\over2\ra^2}.
\eqname\DF
$$
We assume that $f_0(Q)=0$ for $Q<0$.  The luminosity density of these
models
$$\eqalign{
j(r)&=2^{3/2-\beta_0}\pi^{3/2}r^{-2\beta_0}
\left(1+{r^2\over\ra^2}\right)^{-(1-\beta_0)}
{\Gamma(1-\beta_0)\over\Gamma({3\over2}-\beta_0)}\times\cr
&\qquad\int_0^{\psi(r)}f_0(Q)(\psi-Q)^{1/2-\beta_0}\,\d Q,\cr
}
\eqname\lumdens
$$
where $\psi(r)$ is related to the gravitational potential $\Phi(r)$
through $\psi(r)\equiv-\Phi(r)$.  This demonstrates that the function
$f_0(Q)$ is (implicitly) determined once the luminosity density $j(r)$
and mass distribution are specified.

More generally, the non-trivial velocity moments of these models are
given by
$$[v_r^{2k}v_\phi^{2l}v_\theta^{2m}](r)=j\={v_r^{2k}v_\phi^{2l}v_\phi^{2m}}(r)
 = C_{klm}(r)V_{k+l+m}(r),
\eqname\velmom
$$
where
$$V_n(r)\equiv\int_0^{\psi(r)} f_0(Q)(\psi-Q)^{n-\beta_0+1/2}\,\d Q,
\eqname\Vn
$$
and the coefficients
$$\eqalign{
C_{klm}&=2^{k+l+m+1/2-\beta_0}r^{-2\beta_0}
\left(1+{r^2\over\ra^2}\right)^{-(l+m+1-\beta_0)}\times\cr
& \qquad B\left(k+{1\over2},l+m+1-\beta_0\right)
B\left({l+1\over2},{n+1\over2}\right).\cr
}
\eqname\velmomc
$$
This helps to clarify the meaning of the parameters $\beta$ and~$\ra$:
from \ref{velmom} and~\ref{velmomc}, the anisotropy
$$\beta(r)\equiv 1-{\sigma_\phi^2\over\sigma_r^2}=
{ r^2+\beta_0\ra^2\over r^2+\ra^2},
$$
which tends to $\beta_0$ as $r\rightarrow0$, and to 1 for $r\gg\ra$.

\subsection{From $\Upsilon(r)$ to VPs}

It is straightforward to extend the method of section~3 to calculate
the higher-order moments~$\ref{velmom}$ and reconstruct the VPs from
them.  Our method avoids the difficult problem of obtaining $f_0(Q)$
explicitly from~\ref{lumdens}.  It closely follows the approach of
Magorrian \& Binney (1994).

Integrating~\ref{Vn} by parts gives
$$V_n(r)=(n-\beta_0+1/2)\int_r^\infty V_{n-1}(r'){GM(r')\over r'^2}\,\d r',
$$
which is the higher-order analogue of the Jeans' equation~\ref{jeans}.
Having $V_0$ from $j(r)$, it allows the full set of $V_n$ up to any
arbitrary order to be calculated.  The moments
$[v_r^{2k}v_\phi^{2l}v_\theta^{2m}](r)$ then follow from \ref{velmom}
and~\ref{velmomc}.  Projecting the $n^{\rm th}$-order moments along
lines of sight and convolving with seeing as in Section~2 yields the
classical moments of the VPs $\L(R;\vp)$, namely
$$[\vp^n](R)\equiv I\={\vp^n}(R)\equiv I(R)
\int_{-\infty}^\infty \L(R;\vp)\vp^n\,\d\vp.
$$
Each VP can be reconstructed from its moments $\={\vp^n}$ using a
Gram-Charlier series of type~A (Kendall \& Stuart~1943):
$$\L(\vp)={\alpha(\wgc)\over\sigmagh}
\sum_{i=0}^m d_i\hermone_i(\wgc),
\eqname\GCser
$$
where $\wgc\equiv\vp/\sigmagc$,
$\alpha(x)\equiv\exp(-x^2/2)/\sqrt{2\pi}$ is the standard Gaussian,
the $\hermone_i$ are Hermite Polynomials of the first kind (e.g.,
Abramowitz \& Stegun\ 1965), and the coefficients
$$d_i(\sigmagc)\equiv \int_{-\infty}^\infty
 \L(\vp)\hermone_i(\wgc)\d\vp
$$
are just linear combinations of the $\={\vp^n}$.  Kendall \& Stuart
(1943) give conditions sufficient for convergence.  For realistic VPs
to be well-approximated using a small number of $d_i$, the expansion
variable $\sigmagc$ should be chosen so that
$\sigmagc^2\sim\={\vp^2}$.

As in \S3.1, many of the coefficients involved in the numerical
evalution of the $V_n$, $\={\vp^n}$ and $d_i$ depend only on
$\beta_0$, $\ra$ and $j(r)$ and therefore need only be calculated
once, significantly speeding up the calculation of the $d_i(R)$ from
$\Upsilon(r)$.

\subsection{Comparing model and observed VPs}

Since VPs are usually close to Gaussian, observed VPs are usually
reported by giving the first few terms of their Gauss--Hermite
expansion (van der Marel \& Franx~1993; see also Gerhard~1993):
$$\L(\vp)={\gammagh\alpha(\wgh)\over\sigmagh}
\sum_{i=0}^{\infty}h_{2i}H_{2i}(\wgh),
$$
where $\wgh\equiv \vp/\sigmagh$, and
the $H_i$ are Hermite polynomials of the second kind.  There is a
different set of Gauss--Hermite coefficients $(h_0,h_2,h_4,\ldots)$
for each choice of $(\gammagh,\sigmagh)$, but observers fix the latter
to be the line-strength and dispersion of the best-fitting Gaussian to
the VP.  Then $h_0=1$, $h_2=0$ and the lowest-order deviation of the
VP from the best-fitting Gaussian is measured by $h_4$ (van der Marel
\& Franx~1993).

While the most straightforward course of action would be to compare
models and observations using this choice of $(\gammagh,\sigmagh)$,
this has the minor disadvantage that it can only give model
predictions $\hat h_i$ at the same radii as the observations.  To
remedy this, we instead choose $\gammagh=1$, and $\sigmagh(R)$
equal to the smoothly varying dispersion of the best-fit initial
model~\ref{inifit}.  We use the relations (Magorrian \& Binney~1994)
$$\eqalign{
{\p h_l\over\p\gammagh} & = -{1\over\gammagh}h_l,\cr
{\p h_l\over\p\sigmagh} & ={1\over2\sigmagh}\left(
\sqrt{(l+2)(l+1)}h_{l+2}+h_l-\sqrt{l(l-1)}h_{l-2}\right),\cr
}
\eqname\hadj
$$
to make the small adjustments required to bring the Gauss-Hermite
coefficients of the observed VPs to this choice of
$(\gammagh,\sigmagh(R))$.  Conversely, when presenting our final
results, we will use these relations to return the model VPs in the
observational parametrization $(\gammagh,\sigmagh,h_4,h_6,\ldots)$ by
adjusting $(\gammagh,\sigmagh)$ so that $(\hat h_0,\hat h_2)=(1,0)$.

The Gauss--Hermite coefficients of the model galaxy, $\hat h_i(R)$,
can be obtained from the moments $\={\vp^n}$ using the Gram-Charlier
expansion~\ref{GCser} and some judicious manipulation of integrals of
Hermite polynomials.  Magorrian \& Binney (1994) show that there is a
simple linear relation between the two, namely
$$\hat h_k=\sum_{l=0}^\infty c_{kl}d_l,
$$
and describe how to obtain the matrix $c_{kl}(\gammagh,\sigmagh)$.  Of
course, in practice it is neither possible nor necessary to include
all the terms in these infinite sums.  Limited spectral resolution
together with the effects of template mismatch mean that observations
can only yield sensible information on the $h_i$ up to $i=N_h\simeq4$.
For most of our models we find that the error in the predicted $\hat
h_4$ introduced by truncating the series~\refeq1 at $l=10$ is of order
$0.002$, much less than typical observational uncertainties.  As a
stronger test of our moment-based VP calculations, we have calculated
the VPs of various models by solving eq.~\ref{lumdens} numerically and
integrating directly over the DF.  The $h_4$ coefficients calculated
this way differ from the moment-based results by less than $0.002$.

To first order, errors in the $h_i$ are uncorrelated (van der Marel \&
Franx~1993).  Assuming that these errors are Gaussian, the likelihood
of the trial $\Upsilon(r)$ is given by
$$\pr(D|\Upsilon,\beta_0,\ra)=\exp\left(-{1\over2}\chi_h^2\right)
$$
with
$$\chi^2_h[\Upsilon,\beta_0,\ra]
=\sum_{i=1}^{N}\sum_{j=0}^{N_h} \left(\hat
h_j(R_i)-h_j(R_i)\over \Delta h_j(R_i)\right)^2.
\eqname\chisqVP
$$
In practice we truncate this series at $j=2$, since the unknown
flattening of the galaxy affects its $h_4$ profile (i.e., we calculate
the full VPs, but fit only to $\sigma$).  Our method can easily be
modified to deal with other parametrizations of VPs, including
so-called ``non-parametric'' parametrizations.  In most cases,
however, a proper treatment of the correlations between the parameters
is likely to lead to a significant increase in the complexity of the
expression for the likelihood.

\subsection{Including discrete tracers of the VPs}

The rapid fall-off in the surface brightness of galaxies means that it
is very difficult to use stellar absorption lines to measure VP shapes
reliably far from the galaxy centre.  Fortunately, in any galaxy there
will be many stars that have evolved into planetary nebulae (PNe).
These objects emit a significant fraction of their light in the O [III]
line at 5007\AA, making it easy to measure their line-of-sight
velocities.  If we make the reasonable assumption that PNe are drawn
from the same DF as the stars, then the probability that a PN at
projected radius $R$ will have line-of-sight velocity between $\vp$
and $\vp+\d\vp$ is just $\L(R;\vp)\,\d\vp$.  Therefore the likelihood
of measuring $N_{\rm PN}$ PNe, the $i^{\rm th}$ at radius $R_i$ with
line-of-sight velocity $v_i$ and error $\Delta v_i$,
$$\pr(D_{\rm PN}|\Upsilon,\beta_0,\ra)\propto {\prod_{i=1}^{N_{\rm
PN}}\L(R_i;v_{i})\Delta v_{i}},
$$
assuming that the observational uncertainties $\Delta
v_{i}\ll\sigma(R_i)$.
Then, using the results from \S4.2 above, the combined likelihood of
the PNe and VP data is
$$\eqalign{
\pr(D|\Upsilon,\beta_0,\ra) & = \pr(D_{\rm VP}|\Upsilon,\beta_0,\ra)
\pr(D_{\rm PN}|\Upsilon,\beta_0,\ra)\cr
& \propto \exp\left(-{1\over2}\chi_h^2\right)
\prod_{i=1}^{N_{\rm PN}} \L(R_i,v_i)\Delta v_i.\cr
}
\eqname\likeVPPN
$$
Since the shape of $\L(R,v)$ is affected by the galaxy's unknown
flattening, in practice we simply ignore the deviations of the model's
predicted $\L(R,v)$ from their best-fit Gaussians, consistent with our
treatment of $h_4$ measurements above.

\subsection{Summary}

In summary then, the procedure we use to obtain the range of
$\Upsilon(r)$ consistent with observed VPs is as follows:
\item{(1)} Fit an initial guess of the form~\ref{inifit} to the
observed dispersions~$\sigma_i$.

\item{(2)} Use~\ref{hadj} to adjust the observed Gauss--Hermite
coefficients $h_i$ to $\gammagh=1$ and $\sigmagh(R)$ equal to the
dispersion profile of this smooth initial fit.

\item{(3)} Iterate the Metropolis algorithm as in \S3.2, but
including higher-order (typically up to $10^{\rm th}$-order) moments
(\S4.1) rather than just the second-order moments (\S3.1).  Use the
likelihood \ref{likeVPPN} instead of~\ref{likesig}.

\item{(4)} When the Metropolis algorithm is finished, calculate the
confidence limits on the model $\Upsilon(r)$, and the corresponding
$\hat h_i$.  Convert the latter back to the observational
parametrization $(\gammagh,\sigmagh,h_4,\ldots)$ using~\ref{hadj}.

\noindent
If we somehow knew that the galaxy was close to spherical, we could
constrain its anisotropy by constructing models with a variety of
$\beta_0$ and $\ra$ and using equation~\ref{comparemodel} to find the
range of $(\beta_0,\ra)$ preferred by the available data.  In
practice, however, we construct models for just a few choices of
$(\beta_0,\ra)$, compare the predicted $\sigma$ and $h_4$ profiles
with the observations and use the results of Appendix~A to estimate
the allowed range of flattening or anisotropy in the real galaxy.

\begintable*{1}
\caption{{\bf Table 1.} Sources of data used.}
\halign{\hfil#  &   
   \quad\hfil # &  
   \quad\hfil # &  
   \quad\hfil # &  
        \quad # &  
        \quad # &  
        \quad # &  
        \quad # &  
        \quad # &  
        \quad # &  
        \quad #    
	\cr        
         Name & Dist. & $B_{\rm T}$ & $\log L_B$ & 
         $B-V$ & $V-I$ & $r_{\rm eff}$ &
         Phot.  & Kin. & $r_{\rm max}$ & $v_c(r_{\rm eff})$ \cr
            & (Mpc) &    &  $(L_\odot)$  &       &       & (arcsec) & 
         Source & Source & (arcsec) & $(\kms)$ \cr
(1) & (2) & (3) & \hfil(4)\quad & \hfil(5)\quad & \hfil(6)\hfil\quad & \quad(7)\hfil & \hfil(8)\hfil\quad &
\hfil(9)\qquad\hfil&\hfil(10)\qquad&\hfil(11)\quad\cr
\noalign{\vskip2pt}\noalign{\hrule}\noalign{\vskip2pt}
M32 & 0.8 & 8.76 &  8.49&0.89 & 1.14 & \hfil40\qquad 
& K87 & vdM94 & \hfil 22\quad &\hfil$  75\pm 15 $\cr
 & &  &  & & & &  & NF86 (PNe)&\hfil 271\quad & \cr
NGC 1379 & 17.9 & 11.66 & 10.03&0.91 & 1.19 & \hfil24\qquad 
& FIH89 & GCBZL98 & \hfil 61\quad &\hfil$ 167\pm  5 $\cr
NGC 1400 & 21.5 & 11.62 & 10.21&1.03 & 1.37 & \hfil38\qquad 
& L85 & BBBDDSSZZ94 & \hfil 45\quad &\hfil$ 268\pm 20 $\cr
NGC 2434 & 20.0 & 11.43 & 10.22&0.91 & 1.18 & \hfil24\qquad 
& CD94 & CZMDQ95 & \hfil 62\quad &\hfil$ 296\pm 17 $\cr
NGC 3379 & 9.9 & 10.43 & 10.01&0.99 & 1.23 & \hfil35\qquad 
& PDIDC90 & SS99 & \hfil 78\quad &\hfil$ 262\pm 14 $\cr
 & &  &  & & & &  & CJD93 (PNe)&\hfil 209\quad & \cr
NGC 4434 & 15.3 & 12.83 &  9.43&-- & -- & \hfil19\qquad 
& CCR90 & SP97 & \hfil 17\quad &\hfil$ 142\pm 18 $\cr
NGC 4464 & 15.3 & 13.61 &  9.12&0.89 & 1.19 & \hfil6\qquad 
& CCR90 & SF97 & \hfil 17\quad &\hfil$ 165\pm 13 $\cr
NGC 4467 & 15.3 & 14.81 &  8.64&-- & -- & \hfil10\qquad 
& -- & BN90 & \hfil 9\quad &\hfil$  98\pm 10 $\cr
NGC 4472 & 15.3 & 9.32 & 10.83&0.99 & 1.26 & \hfil105\qquad 
& PDIDC90 & FIF95${}^*$ & \hfil 140\quad &\hfil$ 419\pm 23 $\cr
NGC 4478 & 15.3 & 12.14 &  9.71&0.90 & 1.19 & \hfil14\qquad 
& CCR90 & DEFIS83 & \hfil 27\quad &\hfil$ 220\pm 13 $\cr
M87 & 15.3 & 9.52 & 10.75&0.98 & 1.27 & \hfil105\qquad 
& PDIDC90 & ST96 & \hfil 170\quad &\hfil$ 531\pm 17 $\cr
NGC 4494 & 15.3 & 10.69 & 10.29&-- & -- & \hfil45\qquad 
& BM87 & JS89${}^*$ & \hfil 37\quad &\hfil$ 240\pm 25 $\cr
NGC 4552 & 15.3 & 10.84 & 10.23&0.97 & 1.20 & \hfil30\qquad 
& MM94 & SP97 & \hfil 58\quad &\hfil$ 330\pm 17 $\cr
NGC 5846 & 26.0 & 10.67 & 10.75&-- & -- & \hfil63\qquad 
& MM94 & ST96 & \hfil 145\quad &\hfil$ 418\pm 20 $\cr
NGC 6407 & 60.0 & 12.90 & 10.59&1.12 & 1.38 & \hfil33\qquad 
& CD94 & CD94${}^*$ & \hfil 43\quad &\hfil$ 440\pm 32 $\cr
NGC 7192 & 34.0 & 12.04 & 10.44&0.97 & 1.24 & \hfil28\qquad 
& CD94 & CD94${}^*$ & \hfil 43\quad &\hfil$ 308\pm 23 $\cr
NGC 7507 & 21.0 & 11.15 & 10.38&1.00 & 1.28 & \hfil31\qquad 
& FIH89 & BBBDDSSZZ94 & \hfil 66\quad &\hfil$ 321\pm 16 $\cr
NGC 7796 & 37.0 & 12.28 & 10.42&1.00 & 1.23 & \hfil37\qquad 
& SWJTB91 & BBBDDSSZZ94 & \hfil 37\quad &\hfil$ 371\pm 30 $\cr
}
\tabletext
{Distances (col.~2) are taken (in decreasing order of preference) from
Faber et al.\ (1997), Faber et al.\ (1989) or the Hubble relation
assuming $H_0=80\kms\Mpc^{-1}$.
$B$ magnitudes, $B-V$ colours and effective radii (cols.\ 3, 5 and~7)
are taken from Faber et al.\ (1989).  $V-I$ colours (col.~6) are from
Prugniel et al. (1993) and Poulain \& Nieto (1994).  The photometry used
(col.~8) is supplemented with HST photometry from Faber et al.\ (1997)
when available.  Col.~10 gives the radius of the outermost observed point
of the kinematic source (col.~9).  Our models use velocity dispersions
along one axis (usually the minor if available), except for sources
marked with an asterisk for which we use results from multiple slit
positions.
Col.~(11) gives the 95\% confidence intervals on the circular velocities
of the isotropic models
at one effective radius.
}
\endtable

\section {Application to real galaxies}

In this section we apply the machinery to a number of real galaxies,
listed in Table~1.  The sample consists of galaxies no flatter than E2
on the sky, with published kinematics measured out to of order the
galaxy's effective radius or further.  For most of the galaxies the
available data consist simply of a velocity dispersion profile, but in
a handful of cases measurements of the shapes of the VPs ($h_4$
parameters) are also available.  Two of the galaxies have PNe
radial-velocity measurements: 15 radial velocities for M32, 29 for
NGC~3379.

We obtain the luminosity density $j(r)$ of each galaxy by deprojecting
its spherically averaged surface brightness distribution using the
method described in Magorrian (1999) with $n_r=30$ radial grid points.
Then we apply the machinery of \S4 using the observed kinematics to
constrain the galaxy's mass profile, first under the assumption that
the galaxy is isotropic ($\beta=0$), then assuming that it is mildly
radially anisotropic ($\beta=0.3$).  In each case we carry out $10^5$
iterations of the Metropolis algorithm, including terms up to $l=10$
in equation~\ref{GCser}.  But, given the degeneracy between anisotropy
and flattening identified above, we ignore any measurements of VP
shapes, and fit simply to the velocity dispersion profile (i.e., we
truncate the series~\ref{chisqVP} at $N_h=2$), and merely predict the
$h_4$ profile for a spherical galaxy with the assumed anisotropy.

\newdimen\figwid\figwid=0.66\hsize
\beginfigure*{4}
\hfil
            \psfig{file=M32.ps,width=\figwid}\hfil
            \psfig{file=n1379.ps,width=\figwid}\hfil
            \psfig{file=n1400.ps,width=\figwid}\hfil

\bigskip
            \hfil
            \psfig{file=n2434.ps,width=\figwid}\hfil
            \psfig{file=n3379.ps,width=\figwid}\hfil
            \psfig{file=n4434.ps,width=\figwid}\hfil
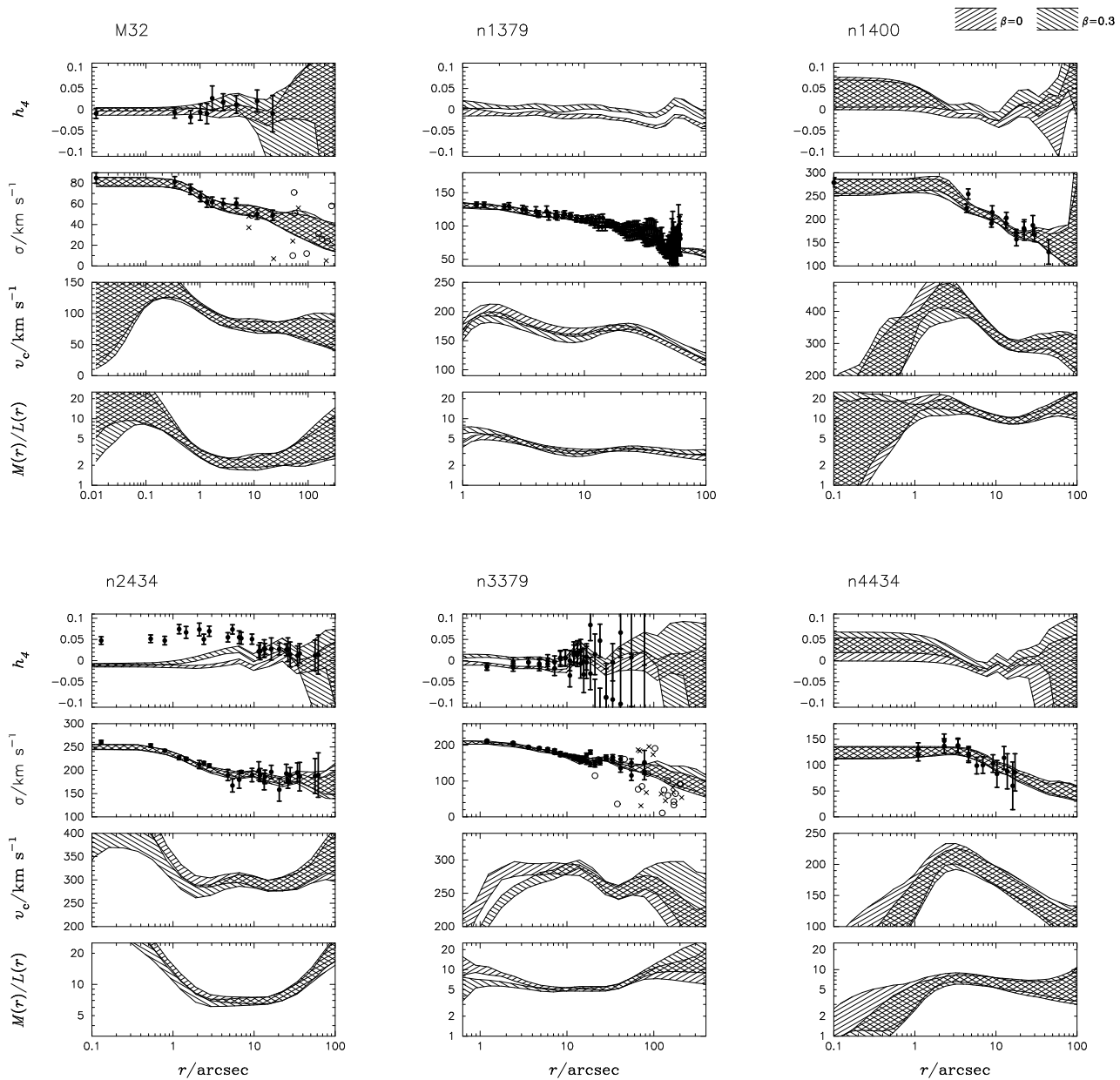
\caption{{\bf Figure 3.}  Model fits for mass-to-light ratios,
circular velocities and projected kinematics for the galaxies in our
sample.  The hatched regions give the 95\% confidence bands for each
quantity.  For each galaxy an isotropic ($\beta=0$) and a mildly
radially anisotropic ($\beta=0.3$) model is constructed by varying the
local mass-to-light ratio to fit the observed dispersion profile.  The
$h_4$ profiles plotted are model predictions, not fits.  The PNe
measurements used in our models of M32 and NGC~3379 are plotted as the
crosses ($v<0$) and open circles ($v>0$).}
\endfigure

\beginfigure*{5}
            \hfil
            \psfig{file=n4464.ps,width=\figwid}\hfil
            \psfig{file=n4467.ps,width=\figwid}\hfil
            \psfig{file=n4472.ps,width=\figwid}\hfil

\bigskip
            \hfil
            \psfig{file=n4478.ps,width=\figwid}\hfil
            \psfig{file=M87.ps,width=\figwid}\hfil
            \psfig{file=n4494.ps,width=\figwid}\hfil
\caption{{\bf Figure 3.}  continued...}
\endfigure

\beginfigure*{6}
            \hfil
            \psfig{file=n4552.ps,width=\figwid}\hfil
            \psfig{file=n5846.ps,width=\figwid}\hfil
            \psfig{file=n6407.ps,width=\figwid}\hfil

\bigskip
            \hfil
            \psfig{file=n7192.ps,width=\figwid}\hfil
            \psfig{file=n7507.ps,width=\figwid}\hfil
            \psfig{file=n7796.ps,width=\figwid}\hfil
\caption{{\bf Figure 3.}  continued.}
\endfigure
\subsection{Results for individual galaxies}

The results of this modelling are plotted on Figure~3.  One of the
most obvious features of the results is that each model's outer mass
profile is almost independent of the anisotropy we assume (Efstathiou,
Ellis \& Carter~1980; Gerhard~1993).  Some of the galaxies deserve
more detailed comment.

{\bf M32}\quad Our models use the minor-axis velocity dispersion
profile obtained by van der Marel et al.\ (1994) and the 15 PNe radial
velocities measured by Nolthenius \& Ford (1986).  These data are
insufficient to place any interesting constraints on whether M32 has a
halo or not: the $v_{\rm c}(r)$ and $\Upsilon(r)$ profiles of our
models are both consistent with being flat.  The results plotted are
for models where we have subtracted a heliocentric systemic velocity
of $-203\kms$ (Tully~1988) from the PNe velocities.  Assuming systemic
velocities of $-195\kms$ or $-185\kms$ (Nolthenius \& Ford~1986)
yields almost identical results.

{\bf NGC 2434}\qquad As mentioned in \S2, Rix et al.\ (1997) have used
Scharzschild's method to construct spherical models of this galaxy.
Their models required a dark halo with $v_{\rm c}\simeq300\kms$ at
around $r_{\rm eff}$.  Our results are in very good agreement.  We
re-iterate, however, that if this galaxy were flattened along the line
of sight these spherical models would underestimate the true circular
velocity.

{\bf NGC 3379}\quad Our models use the minor-axis dispersion profile
measured by Statler \& Smecker-Hane (1999), and the 29 PNe radial
velocities from Ciardullo, Jacoby \& Dejonghe (1993).  We have
subtracted a systemic velocity of $900\kms$ from the PNe velocities
for the model plotted in Fig.~3.  The results are almost identical if
we instead assume a systemic velocity of $916\kms$, equal to the mean
recession velocity of the PNe sample, or if we use the $881\kms$
quoted in Tully (1988).  Statler \& Smecker-Hane (1999) discovered
that this galaxy has a kink in its dispersion profile at a radius of
$\sim15$ arcsec.  Our models fit the data beyond the kink by having
$\Upsilon(r)$ increase by a factor of three between 15 arcsec and 60
arcsec, then becoming further out in order to fit the PNe radial
velocities.  An alternative explanation is that NGC~3379 may be a
face-on S0 (e.g., Capaccioli et al.\ 1991), or a weakly triaxial
system viewed face on (Statler \& Smecker-Hane~1999).  We note that
the the evidence for positive $h_4$ in this galaxy could be due either
to radial anisotropy (Gebhardt et al.\ 2000) or to the galaxy being
flattened system viewed face on (Section~2).

{\bf NGC 4472}\quad Brighenti \& Mathews' (1997) modelling of the
X-ray emission from this galaxy indicates a mass of about
$3\times10^{11}M_\odot$ within $7.8\kpc$ (scaling to our assumed
distance of $15.3\Mpc$), which corresponds to $v_{\rm
c}\simeq410\kms$.  Similar results are found by Irwin \& Sarazin
(1996).  This is in good agreement with our $v_{\rm c}=(419\pm11)\kms$
at $1r_{\rm eff}=7.8\kpc$.

{\bf NGC 4486}\quad Nulsen \& B\"ohringer (1995) find that the X-ray
data for M87 are consistent with a mass per unit length of
$3.6\times10^{10}M_\odot \kpc^{-1}$, or $v_{\rm c}=390\kms$.
Similarly, Tsai (1993) finds a mass $2.8\times10^{11}M_\odot$ enclosed
within a radius $7.8\kpc$, also giving $v_{\rm c}=390\kms$.  Using the
velocity dispersion measurements of Sembach \& Tonry (1996), our
models find $v_{\rm c}=(530\pm9)\kms$, larger by 30\%.  Part of this
discrepancy is due to the fact that Sembach \& Tonry's measurements
are systematically offset by 10 to 15\% (Sembach, private
communication).  Comparison with the measurements of Sargent et al.\
(1978), Bender, Saglia \& Gerhard (1994) and van der Marel (1994)
confirms that Sembach \& Tonry's measurements are systematically
overestimated by about 15\%.

\smallskip\noindent
Taking our models at face value, five of the galaxies in the sample
(M87, NGC 2434, 4472, 5846 and 7796) show evidence for a dark halo
(flat rotation curve and rising mass-to-light ratio at large radii).
Five (NGC 1379, 4434, 4552, 7192, 7507) show evidence against a dark
halo (constant mass-to-light ratio and falling rotation curve).  It
should be borne in mind, however, that these results depend critically
on our assumption of spherical symmetry.  For example, in our model of
NGC~4552 the rotation curve falls off at large radii, and mass follows
light very closely.  Despite being round on the sky, however, this
galaxy is classified as an S0 in some catalogues.  If its light
distribution did indeed become significantly flattened at large radii,
then it would have to have a rising mass-to-light ratio profile, and
possibly a flat rotation curve.

\def\R{{\cal R}}
The remaining eight galaxies (M32, NGC~1400, 3379, 4464, 4467, 4478,
4494 and 6407) are indeterminate, with both a flat mass-to-light ratio
and a flat rotation curve at the outermost available radius.  The most
sensible way of distinguishing between constant-$\Upsilon$ and
constant-$v_{\rm c}$ models of these galaxies would be by using PNe to
estimate the galaxy's mean luminosity-weighted velocity dispersion
outside the radius of the outermost stellar velocity dispersion.  In
Appendix~B we show that distinguishing between constant-$\Upsilon$ and
constant-$v_{\rm c}$ models with 95\% confidence would require radial
velocities of $\sim100$ such PNe.

\beginfigure*{7}
\centerline{\psfig{file=tfB.ps,width=0.5\hsize}
            \psfig{file=tfI.ps,width=0.5\hsize}}
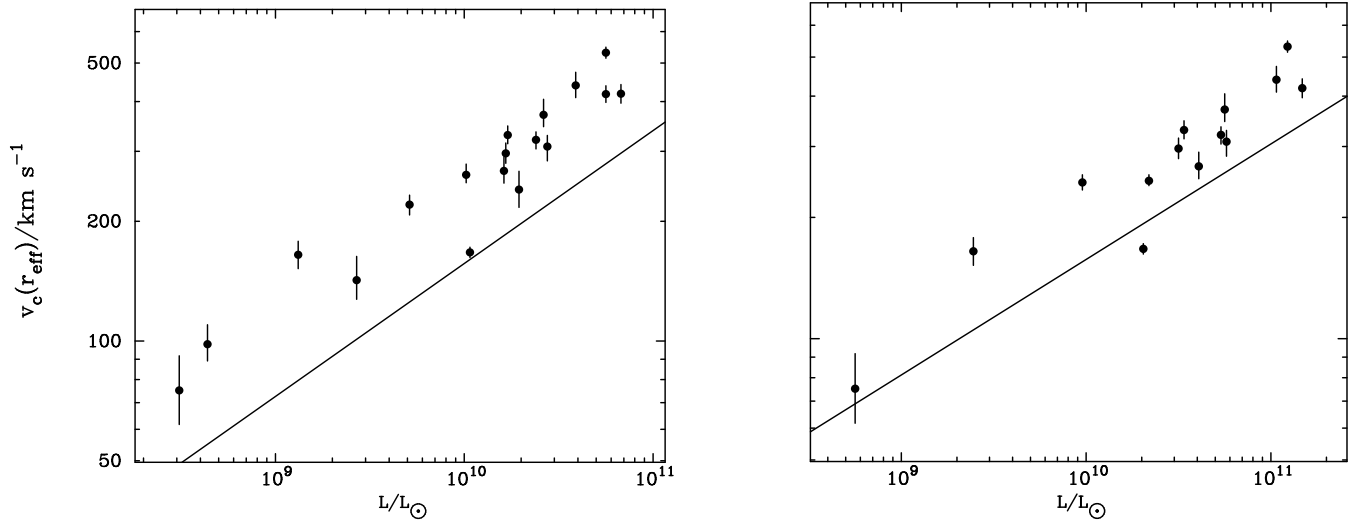
\caption{{\bf Figure 4.} Tully--Fisher relation for our sample of
early-type galaxies.  The points plot the 95\% confidence bounds on
the circular velocity at $1r_{\rm eff}$ against $B$-band (left panel)
and $I$-band (right panel) luminosities.  For comparison, the straight
lines plot the Tully--Fisher relation for spirals as found by Pierce
\& Tully (1992).}
\endfigure

\subsection{The Tully--Fisher relation for ellipticals}

Figure~4 plots the circular velocity derived by our models at $1r_{\rm
eff}$ versus galaxy luminosity.  They lie on a Tully--Fisher relation
parallel to that for spiral galaxies, as was first noted by Franx
(1993).  Our elliptical Tully--Fisher relation is offset from Pierce
\& Tully's (1992) spiral relation in the sense that, for given $v_{\rm
c}$, ellipticals are fainter than spirals by 1.5 mag ($B$ band) or 1.0
mag ($I$ band).  The offset from Mathewson, Ford \& Buchhorn's (1992)
sample is 0.83 mag in~$I$.  Of course, these numbers are
underestimates (overestimates) of the true offset if our ellipticals
are oblate (prolate) galaxies viewed face on.

The scatter in our Tully--Fisher relation for ellipticals is 0.55 mag,
significantly larger than the values $\sim0.2$ mag reported for
samples of nearby spirals.  But the scatter in the real $v_{\rm
c}$--$L$ correlation for ellipticals could well be smaller than 0.55
mag: our galaxy sample is heterogenous, precluding accurate distance
(and therefore luminosity) estimates, and we are unable to correct our
circular velocities for the unknown flattening of our galaxies, a
problem that is usually unavoidable when using elliptical galaxies.

Our results differ from those of Neistein et al.\ (1999), who looked
for a Tully--Fisher relation in a sample of nearby S0 galaxies, using
the method of surface brightness fluctuations to obtain careful
distance estimates, and obtaining circular velocities by deprojecting
the galaxies' observed stellar kinematics.  They found a relation with
an intrinsic scatter of 0.7 mag ($I$ band), with S0s of given $v_{\rm
c}$ only 0.5 mag fainter than corresponding type-Sc spirals.  This is
less than expected if S0s are former spirals with truncated star
formation histories.  Our results can only be brought into broad
agreement with theirs if we assume that our round ellipticals are
mostly face-on prolate systems.  Their method, however, also requires
assumptions about the intrinsic shapes of their galaxies, although it
is unclear whether this could explain the worryingly small 0.5 mag
offset between S0s and spirals.

\section {Conclusions}

We have determined the outer mass profiles in a sample of 18 round
galaxies, under the assumption that each is spherical and has constant
anisotropy.  The assumption of spherical symmetry is the more
dangerous of the two: if a galaxy is flattened along the line of
sight, our spherical models underestimate its mass, whereas varying
the anisotropy has little effect on the outer mass profile.
Nevertheless, if we assume that a galaxy's anisotropy and the
flattening of its light distribution and of its equipotential surfaces
do not vary strongly with radius then we may sensibly use our
spherical models to check for evidence of a dark halo (i.e., a rise in
mass-to-light ratio with radius).

Our results are somewhat unsurprising.  There is very clear evidence
for massive dark haloes in large ellipticals, whereas the the case for
smaller galaxies is more ambiguous.  Our early-type galaxies follow a
Tully--Fisher relation parallel to that for spiral galaxies, but
offset by at least 0.8 mag ($I$-band).  This is larger than the offset
found by Neistein et al.\ (1999) for their sample of S0 galaxies, but
is reassuringly consistent with what one would expect from passive
stellar evolution.

The degeneracy between anisotropy and shape pointed out in \S2 (see
also Magorrian~1999) means that we are unable to place strong
constraints on the anisotropies of our galaxies.  It is often claimed
that elliptical galaxies are radially anisotropic, but there is
surprisingly little solid evidence for this assertion.  For example,
detailed dynamical models have been constructed for the
intermediate-luminosity ellipticals NGC~2434 (Rix et al.\ 1997),
NGC~3379 (Gebhardt et al.\ 2000) and NGC~6703 (Gerhard et al.\ 1998).
In all cases, the models were required to become radially anisotropic
in order to fit the observed VP shapes, but in all cases it was
unjustifiably assumed that the galaxy was either spherical or
spheroidal -- isotropic models with weak face-on disks can also
provide reasonable fits to the VPs of these galaxies.

Brighter ellipticals suffer less from this degeneracy, since they are
unlikely to be disky.  The strongest evidence for radial anisotropy in
an elliptical galaxy comes from Matthias \& Gerhard's (1999)
investigation of NGC~1600, a boxy E4 galaxy.  Their best-fit model has
an anisotropy $\beta\simeq0.5$ in the equatorial plane and
$\beta\simeq0$ along the symmetry axis.  Since this galaxy is so
strongly flattened in projection, it must be close to edge on,
minimizing the uncertainties in the shape of its intrinsic light
distribution, and, therefore, its mass profile and anisotropy
(provided one makes the reasonable assumption that its mass
distribution is not significantly flatter than its light
distribution).

It is unlikely that one will ever be able to directly place strong
constraints on the intrinsic shapes (and therefore anisotropies and
masses) of round galaxies.  To gain a proper understanding of the
dynamical nature of elliptical galaxies, it would be more sensible to
start with a thorough investigation of the shapes and profiles of the
dark and luminous matter in flattened systems like NGC~1600.  Models
of these galaxies when ``observed'' face-on could then prove to be an
interesting test of the nature of rounder galaxies.

\vfil
\section*{Acknowledgments}
We thank Scott Tremaine for helpful advice, Marijn Franx and John
Hibbard for useful discussions, and Ortwin Gerhard for comments on a
draft of the manuscript.  Financial support was provided by NSERC and
PPARC.  D.R.B.\ participated originally via the Physics Co-op program
of the University of Victoria, Canada.

\def\apj #1 #2{ApJ, #1, #2}
\def\apjs #1 #2{ApJS, #1, #2}
\def\aj #1 #2{AJ, #1, #2}
\def\mn #1 #2{MNRAS, #1, #2}
\def\aa #1 #2{A\&A, #1, #2}
\def\aas #1 #2{A\&AS, #1, #2}
\def\araa #1 #2{ARA\&A, #1, #2}

\appendix

\section{The degeneracy between mass, anisotropy and flattening in
scale-free oblate galaxy models}

\beginfigure{8}
\psfig{file=degen.ps,width=\hsize}
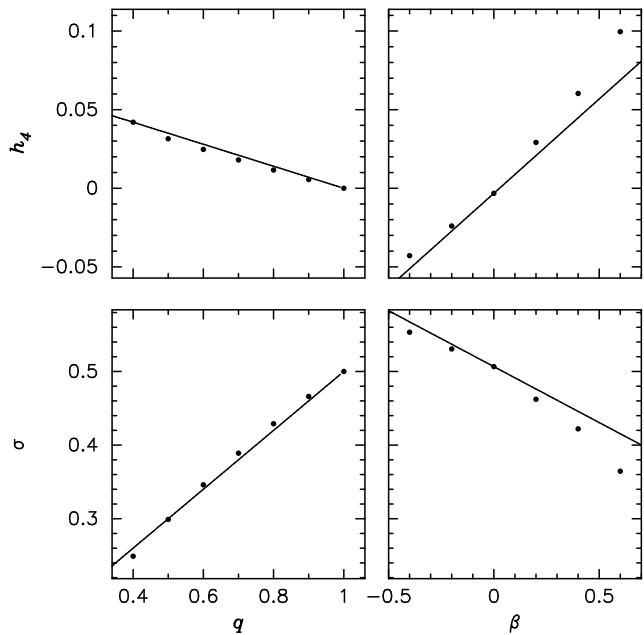
\caption{{\bf Figure~5.} Degeneracy between mass,
anisotropy and flattening in scale-free models.  The panels on the
left show how the Gauss--Hermite coefficients of the VPs of face-on
oblate isotropic models vary with axis ratio.  The panels on the right
show how the VPs of spherical models depend on anisotropy.  All models
have a luminosity density profile $j\sim r^{-4}$ embedded in a
spherical isothermal halo, but similar results are found
for other density profiles and for a Keplerian potential.  The lines
plot the crude fits described in the text.}
\endfigure

\noindent 
Gerhard (1993) has carried out an extensive investigation into how
anisotropy affects the VPs of simple spherical galaxy models.  Here we
use scale-free models to compare the effects of anisotropy with the
effects of flattening the galaxy's light distribution.  For
simplicity we confine our attention to spheroidal models, and ignore
the effects of diskiness, boxiness or a radially varying axis ratio.

Our models have power-law density profiles
$$\nu(R,z)\propto\left(R^2+{z^2\over q^2}\right)^{-\alpha/2},
$$
with axis ratio $q$, and either Kepler $\Phi(r)=-GM/r$ or logarithmic
$\Phi(r)={1\over2}v_{\rm c}^2\ln r$ potentials.  We construct two
orthogonal classes of models: spherical, anisotropic ($q=1$,
$\beta\ne0$) and flattened, isotropic ($q\ne1$, $\beta=0$), viewed
face-on.

\def\vs{v_{\rm s}}\def\phis{\phi_{\rm s}}
We calculate the VPs of the flattened, isotropic models using the
method described in Magorrian \& Binney (1994).
For the spherical anisotropic models we solve equation~\ref{DF} for
$f_0(Q)$ and calculate the VPs by direct integration.  This gives
results that are in very good agreement with the moment-based method
of \S4.
\crap{
For the spherical
anisotropic models we could use the method of \S4 to calculate their
VPs.  As an independent check of that method, however, we also solve
equation~\ref{DF} for $f_0(Q)$, and calculate the VPs by direct
integration,
$$\L(R;\vp)=\int\!\d z\!\int\!\d v_R\,\d v_\phi\,L^{-2\beta_0}f_0(Q).
$$
Here we have chosen a cylindrical co-ordinate system with the $z$-axis
along the line of sight (i.e., $v_z=\vp$).  For the numerical
evaluation of the outer, line-of-sight, integral we change variable
from $z$ to~$u$, where $z=R\sinh u$.  The evaluation of the inner
$(v_R,v_\phi)$ integral is more complicated.  We transform to polar
co-ordinates $(\vs,\phis)$ defined through $v_R=\vs\sin\phis$ and
$v_\phi=\vs\cos\phis$.  The resulting $(\vs,\phis)$ integral can then
be evaluated directly for models with $\beta_0\le0$.  Models with
$\beta_0>0$, however, have an awkward singularity at $L=0$.  For these
models we change variable from $\phis$ to
$t\equiv\ln\tan{1\over2}\phis$.  The VPs calculated using the method
of \S4 agree well with the VPs calculated by this direct method:
the difference in the $h_4$ coefficients predicted using the two
methods is around $10^{-3}$ when moments up to $m=10$ are included in
equation~\ref{GCser}, the agreement improving as $m$ is increased.
}
Figure~5 shows the Gauss--Hermite coefficients of the VPs of models
with $\alpha=4$ in spherical isothermal halos with $v_{\rm c}=1$.
Changing the axis ratio of one of the isotropic models by an amount
$\Delta q$, the Gauss--Hermite cofficients of its VPs change by
$$\eqalign{
{\Delta\sigma\over\sigma}& \simeq0.8\Delta q\cr
\Delta h_4&\simeq-0.07\Delta q.\cr
}
$$
On the other hand, on increasing a spherical galaxy's anisotropy by an
amount $\Delta\beta$, its VPs change by
$$\eqalign{
{\Delta\sigma\over\sigma}&\simeq-0.3\Delta\beta\cr
\Delta h_4&\simeq0.12\Delta\beta.\cr
}
$$
We have checked that these expressions hold for models with
$2\lta\alpha\le4$ in both Kepler and halo potentials.  The effects of
even stronger radial anisotropy can be mimicked by making the
flattened galaxy slightly disky -- the expressions in~\refeq2 apply
only for perfectly oblate galaxies.

There are many galaxies whose observed VPs are more centrally peaked
(larger $h_4$) than the predictions from simple spherical isotropic
models.  One way of fitting these kinematics is to construct spherical
or spheroidal models with mild radial anisotropy (e.g., Rix et al.\
1997; Gerhard et al.\ 1998; Gebhardt et al.\ 2000; Saglia et al.\
2000).  An alternative is to construct flattened models that are
closer to isotropic, possibly with weak, face-on stellar disks.  It is
reasonable to invoke such disks to explain the kinematics of
intermediate-luminosity ellipticals such as NGC~2434, NGC~3379 and
NGC~6703, since many galaxies of similar luminosity are observed to be
isotropic and disky when viewed edge on.  A giant elliptical like
NGC~1399, however, is much less likely to have a stellar disk, meaning
that the degeneracy between anisotropy and shape is more strongly
constrained to lie within the limits set by \refeq2 and~\refeq1.

\section{Distinguishing between models with dark halos and models with
constant mass-to-light ratios}

Galaxies become very faint at large radii, meaning that
absorption-line spectroscopy can be used to measure stellar velocity
dispersions only out to some maximum radius $r_1\sim r_{\rm eff}$.
One way of probing a galaxy's mass distribution at radii larger
than~$r_1$ is to obtain radial velocities of a large number of PNe.
Our aim in this appendix is to calculate the number of PNe that would
be required to distinguish between constant-$\Upsilon$ and
constant-$v_{\rm c}$ models at given confidence level.  We assume that
the anisotropy and intrinsic shape of each galaxy do not vary
strongly with radius (\S2).

Let us write $\hat\sigma_1$ for the galaxy's projected velocity
dispersion at $r_1$ and $\hat\sigma_2$ for the projected
luminosity-weighted velocity dispersion averaged over radii $r>r_1$.
The former is estimated by absorption-line spectroscopy, the latter
from the PNe radial velocities.  For $r_1\sim r_{\rm eff}$, we find
that constant-$\Upsilon$ models of our galaxies predict a ratio
$\hat\R\equiv\hat\sigma_2/\hat\sigma_1\simeq0.7$, whereas
constant-$v_{\rm c}$ models of course predict $\hat\R=1$.  We write
$M_\Upsilon$ for the model with constant mass-to-light ratio
($\hat\R\simeq0.7$) and $M_{\rm halo}$ for the model with a halo
($\hat\R=1$).

The available data $D$ consist of an estimate
$\sigma_1\pm\Delta\sigma_1$ of $\hat\sigma_1$, and radial velocity
measurements $v_1\ldots v_N$ of the $N$ PNe with radii $r>r_1$.  We
assume that $\pr(\hat\sigma_1|D)$ is Gaussian with mean $\sigma_1$ and
dispersion $\Delta\sigma_1$.  The radial velocity measurements provide
an estimate $\sigma_2^2\equiv\sum v_i^2/N-(\Delta v)^2$ of
$\hat\sigma_2^2$, where $\Delta v$ is a typical error in the measured
$v_i$.  By the central limit theorem, when $N$ is large
$\pr(\hat\sigma_2|D)$ may be approximated by a Gaussian with mean
$\sigma_2$ and dispersion $\Delta\sigma_2\equiv\sigma_2[(1+(\Delta
v/\sigma_2)^2)/2N]^{1/2}$.  Therefore
$$\eqalign{
&\pr(\sigma_1\sigma_2|\hat\sigma_1\hat\sigma_2)\,\d\sigma_1\d\sigma_2=
{1\over2\pi}{\d\sigma_1\d\sigma_2\over\Delta\sigma_1\Delta\sigma_2}\hfill\cr
&\hfill\quad\times
\exp\left[-{1\over2}\left(\sigma_1-\hat\sigma_1\over\Delta\sigma_1\right)^2
-{1\over2}\left(\sigma_2-\hat\sigma_2\over\Delta\sigma_2\right)^2\right].\cr
}
$$
Setting $\hat\sigma_2=\hat\R\hat\sigma_1$, $\sigma_2=\R\sigma_1$ and
integrating over $\sigma_1$ yields
$$\eqalign{
\pr(\R|\hat\R)\d\R & \simeq{1\over\sqrt{2\pi}}{\d\R\over\R(f_1^2+f_2^2)^{1/2}}
\exp\left[-{1\over2}{(\R-\hat\R)^2\over\R^2(f_1^2+f_2^2)}\right],\cr
}$$
where $f_i\equiv\Delta\sigma_i/\sigma_i$ are the fractional errors on the
estimates of $\hat\sigma_1$ and $\hat\sigma_2$.

For an observed dispersion ratio $\R$, Bayes' theorem gives the
relative odds of $M_{\rm halo}$ over $M_\Upsilon$ as
$${\pr(M_{\rm halo}|\R)\over\pr(M_\Upsilon|\R)}
={\pr(\R|M_{\rm halo})\over\pr(\R|M_\Upsilon)}
{\pr(M_{\rm halo})\over\pr(M_\Upsilon)}.
$$
Having no prior preference for either model, we take $\pr(M_{\rm
halo})=\pr(M_\Upsilon)$.  Now suppose that $M_\Upsilon$ were the
correct model.  Then the fraction of observations that will correctly
choose this model over $M_{\rm halo}$ can be obtained by integrating
$\pr(\R|M_\Upsilon)$ over all $\R$ for which the odds~\refeq1 favour
$M_\Upsilon$.  Performing this integral, we find that one must have
$(f_1^2+f_2^2)^{1/2}\lta0.1$ if this fraction is to be larger than
95\%.  Therefore, distinguishing between constant-$\Upsilon$ and
constant-$v_{\rm c}$ models at the 95\% confidence level requires
first that one have an estimate of the velocity dispersion at $r_1\sim
r_{\rm eff}$ with fractional error $f_1<10\%$.  If this dispersion
could be measured with 5\% accuracy, then one would need to measure
the mean dispersion outside $r_1$ with a fractional error $f_2<8.7\%$,
which could be accomplished using $N\sim70$ PNe radial velocities.

\bye